\documentclass{IEEEtaes}

\usepackage{color,array,amsthm}
\usepackage{graphicx}

\usepackage{arydshln}

\jvol{XX}
\jnum{XX}
\jmonth{XXXXX}
\paper{1234567}
\pubyear{2022}
\doiinfo{TAES.2022.Doi Number}

\usepackage{diagbox}
\usepackage{booktabs}
\usepackage{adjustbox} 
\usepackage[utf8]{inputenc}
\usepackage[table]{xcolor}

\usepackage{amsthm}
\usepackage{amssymb}
\usepackage{amsmath,amsfonts}
\usepackage{algorithmic}
\usepackage{algorithm}
\usepackage{array}

\usepackage[caption=false,font=normalsize,labelfont=sf,textfont=sf]{subfig}
\usepackage[font=small,justification=justified,singlelinecheck=false]{caption}
\usepackage{booktabs}
\usepackage{siunitx}
\usepackage{multirow}

\usepackage{threeparttable}

\begin{document}

\title{Generative Spectrum Cartography: Unified Reconstruction and Active Sensing via Diffusion Models}

\author{Yuntong Gu}
\affil{Zhejiang University-University of Illinois Urbana-Champaign Institute (ZJUI), Zhejiang University, Haining, China}

\author{Xiangming Meng}
\member{Member, IEEE}
\affil{Zhejiang University-University of Illinois Urbana-Champaign Institute (ZJUI), Zhejiang University, Haining, China} 

\author{Zhiyuan Lin}
\member{Student Member, IEEE}
\affil{Tsinghua Space Center and Beijing National Research Center for Information Science and Technology, Tsinghua University, Beijing, China}

\author{Sheng Wu}
\member{Member, IEEE}
\affil{Key Laboratory of Universal Wireless Communications, Ministry of Education, School of Information and Communication Engineering, Beijing University of Posts and Telecommunications, Beijing, China}

\author{Linling Kuang}
\member{Member, IEEE}
\affil{Tsinghua Space Center and Beijing National Research Center for Information Science and Technology, Tsinghua University, Beijing, China}

\markboth{GU ET AL.}{GENERATIVE SPECTRUM CARTOGRAPHY}
\maketitle

\begin{abstract}
High-fidelity spectrum cartography is important for spectrum monitoring and wireless situational awareness, especially in satellite-based wide-area sensing scenarios where measurements are sparse, noisy, and often low-bit quantized. In such settings, two coupled challenges arise: accurate reconstruction from severely incomplete measurements and efficient allocation of additional sensing resources under a limited sensing budget. Existing methods usually address these problems separately, and, for reconstruction, they often rely on priors that are insufficiently expressive under sparse and quantized measurements. This paper proposes Generative Spectrum Cartography (GSC), a diffusion-based posterior inference framework for spectrum cartography with uncertainty-aware active sensing. Specifically, spectrum map recovery is formulated as a Bayesian inverse problem under a learned diffusion model  prior, and closed-form posterior mean updates are derived for both linear and quantized measurement models. By embedding these updates into the reverse diffusion process, GSC enables gradient-free and measurement-consistent posterior sampling without relying on computationally costly likelihood-gradient guidance. The resulting posterior samples are further used to estimate spatial uncertainty and to guide diversity-aware selection of additional measurement locations for active sensing. Experiments on simulated electromagnetic maps and a high-fidelity simulated satellite monitoring scenario show that GSC achieves higher PSNR, lower LPIPS, and more efficient sensing than representative baseline methods under sparse, noisy, and low-bit quantized measurements.
\end{abstract}

\begin{IEEEkeywords}
Spectrum Cartography; Generative Models; Inverse Problems; Quantized Measurements; Diffusion Models.

\end{IEEEkeywords}

\section{INTRODUCTION}
\label{sec:introduction}

Electromagnetic spectrum cartography aims to reconstruct the spatial distribution of power spectral density (PSD) over a target region from spatially incomplete measurements \cite{shrestha2022deep,reddy2022spectrum}. Such spatial spectrum information is fundamental to wireless communications, spectrum management and sharing, and radio resource allocation \cite{4699911,8648450}. It is also increasingly relevant to satellite communications and wide-area electromagnetic environment monitoring \cite{article1,9520322}. By characterizing regional spectrum occupancy, propagation conditions, and interference structure, high-fidelity spectrum maps can support network planning, resource allocation, and operational decision-making. Owing to its broad relevance to integrated sensing and communications, dynamic spectrum access, and satellite-based remote monitoring, spectrum cartography has attracted increasing attention in recent years \cite{shrestha2022deep,reddy2022spectrum,10764739,Xu2021}.

In practice, however, acquiring dense and reliable spectrum measurements over a wide area remains difficult, which makes practical spectrum cartography challenging in two coupled aspects. First, reconstructing the underlying PSD field is difficult because the measurements available over a target region are often sparse. In spaceborne wide-area spectrum monitoring, such limitations arise from orbital visibility, coverage or beam geometry, and mission resources \cite{article1}. Existing studies on satellite spectrum monitoring have further noted long data-update cycles, incomplete monitoring data, and additional difficulty in monitoring dynamic uplink transmissions with narrow directional beams \cite{9520322}. Meanwhile, practical measurements may also be affected by non-ideal factors such as measurement noise and low-bit quantization. As a result, recovering the underlying PSD field becomes a severely ill-posed inverse problem rather than a simple spatial interpolation task. Beyond reconstruction accuracy, efficient sensing resource allocation is another critical issue in practical spectrum monitoring systems. After an initial set of sparse measurements is obtained, only a limited number of additional measurements can typically be collected, because each new measurement requires the sensing system to sample a previously unobserved spatial location and consumes sensing time, energy, bandwidth, and other platform resources. The problem is therefore not only to reconstruct the spectrum map from the currently available measurements, but also to determine which unobserved locations should be measured next so that the overall map quality can be improved most effectively under a limited sensing budget.

To address the first challenge, a substantial body of work has studied reconstruction from incomplete measurements from several perspectives. Classical interpolation and statistical estimation methods have been widely used for this purpose \cite{2015ITCCN...1..406D,denkovski2012reliability,bazerque2011group,7952827,6362597}. Representative examples include inverse distance weighting (IDW) \cite{2015ITCCN...1..406D,denkovski2012reliability} and Kriging \cite{6362597}, which reconstruct the field by exploiting local smoothness, spatial correlation, or stationarity assumptions. These methods are appealing for their simplicity and interpretability, but their performance often degrades when the propagation environment is highly non-stationary or when the measurement density becomes very low. Compressive-sensing-based methods impose sparsity in a transform domain and can improve recovery from incomplete data \cite{marin2018compressive,jayawickrama2013improved,5352337,zhang2020spectrum}. Nevertheless, they still rely on hand-crafted structural assumptions and are less effective when the spectrum field exhibits complex multi-source patterns or when the sensing process departs from ideal linear measurement models. More recently, data-driven neural approaches have learned spatial priors directly from electromagnetic map data and have shown improved reconstruction performance over purely model-based baselines \cite{shrestha2022deep,7178572,Xu2021,10906396}. Despite their progress, many such methods remain tied to a specific sampling pattern, measurement model, or quantization rule and may require retraining when these conditions change.

Compared with the extensive effort devoted to accurate reconstruction, the sensing side has received much less attention. Most existing spectrum cartography studies focus on reconstructing a map from a fixed measurement set and do not explicitly use the current reconstruction to guide subsequent measurement selection. As a consequence, information revealed by the current estimate is not fully exploited to guide subsequent measurements.

From a Bayesian perspective \cite{murphy2012mlpp,bishop2006prml}, these two challenges, i.e., high-fidelity spectrum map reconstruction and efficient sensing resource allocation,  can be addressed within a common inference framework. For reconstruction, the objective is to combine the measurement model with a prior over plausible spectrum maps and infer the posterior distribution of the unknown field. For sensing, this posterior distribution provides uncertainty information that can guide future measurements toward locations with higher expected information value. Under this perspective, reconstruction and sensing are not isolated tasks. Instead, they constitute two coupled stages of a unified inference and decision process. The effectiveness of such a framework, however, depends critically on whether the prior is sufficiently expressive and whether posterior inference remains tractable under sparse and quantized measurements.

Recent generative models, particularly diffusion models \cite{song2019generative,ho2020denoising,nichol2021improved,song2020denoising,dou2024diffusion,song2020score} and flow-based models \cite{liu2022flow,lipman2022flow}, offer a promising route to realize this Bayesian perspective, because they learn complex data distributions directly from samples rather than relying on hand-crafted priors. In particular, diffusion models have shown strong capability in inverse problems such as restoration and missing-data recovery under incomplete measurements \cite{zhang2025improving,wang2022zero,kawar2022denoising}. Representative solvers including DPS \cite{chung2023diffusion}, DMPS \cite{meng2022diffusion}, and PiGDM \cite{song2023pseudoinverse} further demonstrate the value of combining a generative prior with a measurement model. Nevertheless, these methods do not directly resolve the spectrum cartography setting considered in this paper. First, because they typically require repeated gradient evaluations during the iterative reverse process, their computational cost can become high. Second, under low-bit quantized measurements, their gradient-based correction mechanisms may be difficult to apply, since the measurement model is nonlinear and may yield unstable or weak gradient information. In addition, existing generative inverse solvers mainly emphasize reconstruction itself, while the use of measurement-conditioned generative inference for active sensing in spectrum cartography remains largely unexplored.

This paper addresses reconstruction and sensing in spectrum cartography jointly from a unified Bayesian perspective and proposes Generative Spectrum Cartography (GSC), a diffusion-based posterior inference framework. We focus on reconstructing a two-dimensional PSD map at a fixed frequency from incomplete measurements, with satellite-based wide-area monitoring serving as the primary motivating application. A pretrained diffusion model is used as a learned prior over plausible spectrum maps. Instead of incorporating measurements through iterative evaluations of the likelihood score function, which can be computationally costly and become unstable under low-bit quantized measurements, we derive analytical measurement-conditioned posterior mean updates under both linear and quantized measurement models and embed these updates into the reverse diffusion process. This yields a gradient-free reconstruction method that incorporates sparse measurements directly into generative posterior inference. Furthermore, using multiple posterior samples generated by the measurement-conditioned reconstruction process, we estimate spatial uncertainty and develop an uncertainty-aware active sensing strategy with diversity-aware selection of measurement locations.

The main contributions of this work are summarized as follows:
\begin{itemize}
    \item \emph{GSC framework for reconstruction and sensing:} We propose Generative Spectrum Cartography (GSC), a posterior inference framework that unifies spectrum map reconstruction and active sensing under a learned diffusion prior. This framework connects posterior-guided reconstruction and sensing-location selection, thereby establishing a closed-loop approach to spectrum cartography under a limited sensing budget.

    \item \emph{Closed-form posterior updates:} We derive analytical measurement-conditioned posterior mean updates for both linear and quantized measurement models. By embedding these updates into the reverse diffusion process, GSC enables gradient-free and measurement-consistent posterior sampling, while avoiding the computationally costly likelihood-gradient guidance used in existing diffusion-based inverse solvers.

    \item \emph{Uncertainty-aware active sensing:} We estimate spatial uncertainty from posterior samples generated by the measurement-conditioned reconstruction process and use it to guide active sensing. To avoid spatially redundant measurements, we further incorporate diversity-aware selection so that additional sensing locations are both informative and well distributed.

    \item \emph{Validation in simulated electromagnetic and satellite monitoring scenarios:} We evaluate GSC on a simulated electromagnetic map dataset and a high-fidelity simulated satellite monitoring scenario using peak signal-to-noise ratio (PSNR) and Learned Perceptual Image Patch Similarity (LPIPS) as the primary fidelity and perceptual metrics. The results show that, under sparse, noisy, and low-bit quantized measurements, GSC achieves improved reconstruction performance and improved sensing efficiency. For example, under 1-bit quantization with a 15\% measurement ratio, GSC improves PSNR from 10.95 dB to 20.98 dB and reduces LPIPS from 0.491 to 0.183 relative to UNN \cite{11127065}.
\end{itemize}

The remainder of this paper is organized as follows. Section~\ref{sec:problem} presents the problem formulation together with the diffusion-model preliminaries. Section~\ref{sec:posterior inference} develops the proposed GSC framework for measurement-conditioned posterior inference in spectrum reconstruction. Section~\ref{sec:uncertainty sensing} introduces the uncertainty-aware active sensing strategy built upon the reconstruction framework. Section~\ref{sec:experiment} provides experimental results on simulated electromagnetic maps and a simulated satellite monitoring scenario. Finally, Section~\ref{sec:conclusion} concludes the paper.

\section{Problem Formulation and Preliminaries}
\label{sec:problem}
\subsection{Problem Formulation}

Following the fixed-frequency setting in \cite{10764739}, we formulate spectrum cartography as an inverse problem of recovering a PSD map $\mathbf{X}_0\in\mathbb{R}^{I\times J}$ from incomplete measurements. Let $\boldsymbol{x}_0=\mathrm{vec}(\mathbf{X}_0)\in\mathbb{R}^{N}$, with $N=IJ$, denote the unknown vectorized PSD map. 

The spatial sampling pattern is described by a diagonal binary sampling operator $\mathbf{H}=\mathrm{diag}(h_1,\ldots,h_N)$, where $h_i\in\{0,1\}$ indicates whether the $i$th grid entry is measured. We denote by $\Omega=\{i:h_i=1\}$ the set of measured indices. The measurement vector is denoted by $\boldsymbol{y}\in\mathbb{R}^{N}$ and written in padded form on the same grid as $\boldsymbol{x}_0$. Let
$
L = |\Omega| = \sum_{i=1}^{N} h_i
$
denote the number of measured locations. We then define the measurement ratio as
$
\rho = \frac{L}{N},
$
which represents the fraction of spatial locations that are measured.

In this paper, we consider two practically relevant measurement scenarios: direct linear measurements and low-bit quantized measurements obtained after analog-to-digital conversion of the linear measurements.

\subsubsection{Linear Measurement Model}

For the linear measurement scenario, we write the forward model as
\begin{equation}
\boldsymbol{y}=\mathbf{H}\boldsymbol{x}_0+\boldsymbol{n},
\label{eq:linear-obs-model}
\end{equation}
where $\boldsymbol{n}\sim\mathcal{N}(\mathbf{0},\sigma_y^2\mathbf{I})$ is additive Gaussian measurement noise. For each $i\in\Omega$, the corresponding measurement satisfies $y_i=x_{0,i}+n_i$. For $i\notin\Omega$, we have $y_i=n_i$. The task is to infer the complete map $\boldsymbol{x}_0$ from sparse and noisy measurements $\boldsymbol{y}$.

\subsubsection{Quantized Measurement Model}

In practical spectrum monitoring systems, the measured analog values may be quantized by low-resolution analog-to-digital converters (ADCs) before transmission or storage \cite{meng2022quantized,zymnis2009compressed,meng2024qcs}. To account for this effect, we extend the linear measurement model in \eqref{eq:linear-obs-model} to the quantized form
\begin{equation}
\boldsymbol{y}=\mathcal{Q}(\mathbf{H}\boldsymbol{x}_0+\boldsymbol{n}),
\label{eq:quant_model}
\end{equation}
where $\boldsymbol{n}\sim\mathcal{N}(\mathbf{0},\sigma_y^2\mathbf{I})$ and $\mathcal{Q}(\cdot)$ denotes an element-wise memoryless scalar quantizer. The quantizer is characterized by decision thresholds
\begin{equation}
-\infty=\tau_0<\tau_1<\cdots<\tau_B=+\infty
\label{eq:quant_thresholds}
\end{equation}
and the corresponding representation levels $\{q_1,\ldots,q_B\}$. For each $i\in\Omega$, the quantized measurement satisfies
\begin{equation}
y_i=q_b
\quad\Longleftrightarrow\quad
\tau_{b-1}<x_{0,i}+n_i\le\tau_b.
\label{eq:quantization_rule}
\end{equation}
This formulation covers general scalar quantization and does not require uniformly spaced thresholds. Unlike the linear model, the quantized model induces a nonlinear and non-Gaussian measurement likelihood, which makes posterior inference significantly more difficult, especially in the low-bit regime \cite{meng2022quantized,meng2024qcs}.

\subsection{Bayesian Posterior Inference Perspective}
\label{sec:bayesian}

From a Bayesian perspective, spectrum reconstruction aims to infer the unknown PSD map $\boldsymbol{x}_0$ from the measurement vector $\boldsymbol{y}$ by combining the measurement model with a prior over plausible maps. The global reconstruction target is the posterior distribution
\begin{equation}
p(\boldsymbol{x}_0 \mid \boldsymbol{y})
\propto
p(\boldsymbol{y} \mid \boldsymbol{x}_0)\,p(\boldsymbol{x}_0),
\label{eq:global_posterior_target}
\end{equation}
where $\propto$ denotes equality up to a normalizing constant independent of $\boldsymbol{x}_0$, $p(\boldsymbol{y} \mid \boldsymbol{x}_0)$ is the likelihood induced by the measurement model, and $p(\boldsymbol{x}_0)$ is a prior distribution over plausible PSD maps.

Equation \eqref{eq:global_posterior_target} shows that reconstruction quality depends on both the measurement model and the prior model. Under sparse and noisy measurements, and especially under nonlinear quantized measurements, the likelihood can be highly ill-conditioned. In such cases, accurate reconstruction requires a sufficiently expressive prior. Classical hand-crafted priors, including smoothness assumptions and sparsity priors commonly used in interpolation and compressive sensing, may be inadequate when the spectrum map exhibits complex multi-source spatial structure or when the measurement process departs from ideal linear settings. This motivates the use of a learned generative prior rather than a hand-crafted prior model.

\subsection{Denoising Diffusion Models}
\label{sec:diffusion}

To model the prior term $p(\boldsymbol{x}_0)$ in \eqref{eq:global_posterior_target}, we adopt a pretrained Denoising Diffusion Probabilistic Model (DDPM) \cite{ho2020denoising,song2020denoising}, which provides a data-driven prior beyond hand-crafted smoothness or sparsity assumptions. In the present problem, the clean sample in the diffusion model is exactly the unknown PSD map $\boldsymbol{x}_0$. DDPM represents this prior through a forward noising process and a reverse denoising process.

In the forward diffusion process, the clean sample $\boldsymbol{x}_0$ is gradually perturbed into Gaussian noise according to
\begin{equation}
q(\boldsymbol{x}_t | \boldsymbol{x}_{t-1})
=
\mathcal{N}
\left(
\boldsymbol{x}_t;
\sqrt{\alpha_t}\,\boldsymbol{x}_{t-1},
(1-\alpha_t)\mathbf{I}
\right),
\label{eq:forward_diffusion}
\end{equation}
where $\alpha_t = 1-\beta_t$, $\{\beta_t\}_{t=1}^{T}$ is a prescribed variance schedule. Equivalently, conditioned on $\boldsymbol{x}_0$, the latent variable $\boldsymbol{x}_t$ admits the closed-form marginal
\begin{equation}
q(\boldsymbol{x}_t | \boldsymbol{x}_0)
=
\mathcal{N}
\left(
\boldsymbol{x}_t;
\sqrt{\bar{\alpha}_t}\,\boldsymbol{x}_0,
(1-\bar{\alpha}_t)\mathbf{I}
\right).
\label{eq:forward_marginal}
\end{equation}
where $\bar{\alpha}_t = \prod_{s=1}^{t}\alpha_s$.
As $t\to T$, the distribution of $\boldsymbol{x}_T$ approaches a standard Gaussian distribution.

The reverse process aims to sample from the transition $p(\boldsymbol{x}_{t-1} | \boldsymbol{x}_t)$, which can be expressed as
\begin{equation}
p(\boldsymbol{x}_{t-1} | \boldsymbol{x}_t)
=
\int
q(\boldsymbol{x}_{t-1} | \boldsymbol{x}_t,\boldsymbol{x}_0)\,
p(\boldsymbol{x}_0 | \boldsymbol{x}_t)\,
d\boldsymbol{x}_0.
\label{eq:reverse_marginal_uncond}
\end{equation}
The first term $q(\boldsymbol{x}_{t-1} | \boldsymbol{x}_t,\boldsymbol{x}_0)$ is the exact reverse posterior conditioned on the clean signal, and it remains Gaussian:
\begin{equation}
q(\boldsymbol{x}_{t-1} | \boldsymbol{x}_t,\boldsymbol{x}_0)
=
\mathcal{N}
\left(
\boldsymbol{x}_{t-1};
\tilde{\boldsymbol{\mu}}_t(\boldsymbol{x}_t,\boldsymbol{x}_0),
\tilde{\sigma}_t^2\mathbf{I}
\right),
\label{eq:ddpm_reverse_diffusion}
\end{equation}
where the reverse mean takes the affine form
\begin{equation}
\tilde{\boldsymbol{\mu}}_t(\boldsymbol{x}_t,\boldsymbol{x}_0)
=
a_t\boldsymbol{x}_0+b_t\boldsymbol{x}_t,
\label{eq:true_ddpm_mu}
\end{equation}
with
\begin{equation}
a_t=
\frac{\sqrt{\bar{\alpha}_{t-1}}\beta_t}{1-\bar{\alpha}_t},
\quad
b_t=
\frac{\sqrt{\alpha_t}(1-\bar{\alpha}_{t-1})}{1-\bar{\alpha}_t},
\label{eq:ddpm_ab}
\end{equation}
and
\begin{equation}
\tilde{\sigma}_t^2=
\frac{1-\bar{\alpha}_{t-1}}{1-\bar{\alpha}_t}\beta_t.
\label{eq:ddpm_variance}
\end{equation}
The second term $p(\boldsymbol{x}_0 | \boldsymbol{x}_t)$ is the diffusion-induced conditional distribution of the clean signal. Rather than explicitly handling this distribution during reverse sampling, DDPM trains a denoiser to estimate the clean signal $\boldsymbol{x}_0$ from the noisy state $\boldsymbol{x}_t$. Under the noise prediction parameterization, this estimate, denoted by $\hat{\boldsymbol{x}}_{0|t} \triangleq \mathbb{E}[\boldsymbol{x}_0 | \boldsymbol{x}_t]$, is written as
\begin{equation}
\hat{\boldsymbol{x}}_{0|t}
=
\frac{1}{\sqrt{\bar{\alpha}_t}}
\left(
\boldsymbol{x}_t
-
\sqrt{1-\bar{\alpha}_t}\,
\boldsymbol{\epsilon}_\theta(\boldsymbol{x}_t,t)
\right),
\label{eq:x0_estimate_from_noise}
\end{equation}
where the neural network $\boldsymbol{\epsilon}_\theta(\boldsymbol{x}_t,t)$ is trained to predict the injected noise at diffusion step $t$.
Given $\hat{\boldsymbol{x}}_{0|t}$ in \eqref{eq:x0_estimate_from_noise} and \eqref{eq:ddpm_reverse_diffusion},
using a delta approximation
$
p(\boldsymbol{x}_0 | \boldsymbol{x}_t)
\approx
\delta(\boldsymbol{x}_0-\hat{\boldsymbol{x}}_{0|t}),
\label{eq:dirac_uncond}
$
DDPM obtains the transition probability as
\begin{align}
p(\boldsymbol{x}_{t-1} | \boldsymbol{x}_t)
&\approx
\int
q(\boldsymbol{x}_{t-1} | \boldsymbol{x}_t,\boldsymbol{x}_0)\,
\delta(\boldsymbol{x}_0-\hat{\boldsymbol{x}}_{0|t})\,
d\boldsymbol{x}_0
\nonumber\\
&=
q(\boldsymbol{x}_{t-1} | \boldsymbol{x}_t,\hat{\boldsymbol{x}}_{0|t})
\nonumber\\
&=
\mathcal{N}
\left(
\boldsymbol{x}_{t-1};
\tilde{\boldsymbol{\mu}}_t(\boldsymbol{x}_t,\hat{\boldsymbol{x}}_{0|t}),
\tilde{\sigma}_t^2\mathbf{I}
\right).
\label{eq:reverse_marginal_uncond_approx}
\end{align}
Accordingly, the standard DDPM reverse update takes the form
\begin{equation}
\boldsymbol{x}_{t-1}
=
\tilde{\boldsymbol{\mu}}_t(\boldsymbol{x}_t,\hat{\boldsymbol{x}}_{0|t})
+
\tilde{\sigma}_t\boldsymbol{z},
\quad
\boldsymbol{z}\sim\mathcal{N}(\mathbf{0},\mathbf{I}).
\label{eq:ddpm_final_xt_minus_1}
\end{equation}

\section{Measurement-Conditioned Posterior Inference for Spectrum Reconstruction}
\label{sec:posterior inference}

Given the Bayesian inverse formulation in Section~\ref{sec:problem}-\ref{sec:bayesian}, the global objective of spectrum reconstruction is to infer the posterior distribution $p(\boldsymbol{x}_0 | \boldsymbol{y})$, which characterizes all plausible reconstructions consistent with both the measurement model and the learned prior.

Under the diffusion prior introduced in Section II, however, posterior sampling is performed not directly in the clean-map space but along the latent diffusion chain $\{\boldsymbol{x}_t\}_{t=1}^{T}$, where $\boldsymbol{x}_t\sim p(\boldsymbol{x}_t | \boldsymbol{y})$. Accordingly, the reverse transition becomes
\begin{equation}
p(\boldsymbol{x}_{t-1} | \boldsymbol{y})
=
\int
p(\boldsymbol{x}_{t-1} | \boldsymbol{x}_t,\boldsymbol{y})\,
p(\boldsymbol{x}_t | \boldsymbol{y})\,
d\boldsymbol{x}_t,
\label{eq:posterior_marginal_recursion}
\end{equation}
and the practical task at reverse step $t$ is therefore to characterize the measurement-conditioned transition $p(\boldsymbol{x}_{t-1} | \boldsymbol{x}_t,\boldsymbol{y})$, which can be expressed as
\begin{equation}
p(\boldsymbol{x}_{t-1} | \boldsymbol{x}_t,\boldsymbol{y})
=
\int
p(\boldsymbol{x}_{t-1} | \boldsymbol{x}_t,\boldsymbol{x}_0,\boldsymbol{y})\,
p(\boldsymbol{x}_0 | \boldsymbol{x}_t,\boldsymbol{y})\,
d\boldsymbol{x}_0.
\label{eq:reverse_marginal_cond}
\end{equation}
Since the measurement $\boldsymbol{y}$ is generated from $\boldsymbol{x}_0$ and therefore provides no additional information about $\boldsymbol{x}_{t-1}$ once $(\boldsymbol{x}_t,\boldsymbol{x}_0)$ are fixed, we have
\begin{equation}
p(\boldsymbol{x}_{t-1} | \boldsymbol{x}_t,\boldsymbol{x}_0,\boldsymbol{y})
=
p(\boldsymbol{x}_{t-1} | \boldsymbol{x}_t,\boldsymbol{x}_0)
=
q(\boldsymbol{x}_{t-1} | \boldsymbol{x}_t,\boldsymbol{x}_0),
\label{eq:reverse_conditional_identity}
\end{equation}
which is exactly the same as \eqref{eq:ddpm_reverse_diffusion}.
Therefore, the exact measurement-conditioned reverse transition can be written as
\begin{equation}
p(\boldsymbol{x}_{t-1} | \boldsymbol{x}_t,\boldsymbol{y})
=
\int
q(\boldsymbol{x}_{t-1} | \boldsymbol{x}_t,\boldsymbol{x}_0)\,
p(\boldsymbol{x}_0 | \boldsymbol{x}_t,\boldsymbol{y})\,
d\boldsymbol{x}_0.
\label{eq:reverse_transition_integral}
\end{equation}
Comparing \eqref{eq:reverse_transition_integral} with \eqref{eq:reverse_marginal_uncond}, the only difference from the unconditional case is that  $p(\boldsymbol{x}_0 | \boldsymbol{x}_t)$ is replaced by $p(\boldsymbol{x}_0 |\boldsymbol{x}_t,\boldsymbol{y})$. So we only need to replace the unconditional clean-signal  $\hat{\boldsymbol{x}}_{0\mid t}$ with its measurement-conditioned estimate $\hat{\boldsymbol{x}}_{0\mid t,\boldsymbol{y}} \triangleq \mathbb{E}[\boldsymbol{x}_0 \mid \boldsymbol{x}_t,\boldsymbol{y}]$. As a result, the measurement-condition transition is obtained as 
\begin{align}
p(\boldsymbol{x}_{t-1} | \boldsymbol{x}_t,\boldsymbol{y})
&\approx
\int
q(\boldsymbol{x}_{t-1} | \boldsymbol{x}_t,\boldsymbol{x}_0)\,
\delta(\boldsymbol{x}_0-\hat{\boldsymbol{x}}_{0|t,\boldsymbol{y}})
\,d\boldsymbol{x}_0
\nonumber\\
&=
q(\boldsymbol{x}_{t-1} | \boldsymbol{x}_t,\hat{\boldsymbol{x}}_{0|t,\boldsymbol{y}})
\nonumber\\
&=
\mathcal{N}
\left(
\boldsymbol{x}_{t-1};
\tilde{\boldsymbol{\mu}}_t(\boldsymbol{x}_t,\hat{\boldsymbol{x}}_{0|t,\boldsymbol{y}}),
\tilde{\sigma}_t^2\mathbf{I}
\right),
\label{eq:posterior_reverse_dirac}
\end{align}
where
\begin{equation}
\tilde{\boldsymbol{\mu}}_t(\boldsymbol{x}_t,\hat{\boldsymbol{x}}_{0|t,\boldsymbol{y}})
=
a_t\,\hat{\boldsymbol{x}}_{0|t,\boldsymbol{y}}
+
b_t\boldsymbol{x}_t.
\label{eq:posterior_reverse_mean_def}
\end{equation}
The measurement-conditioned reverse step can be written explicitly as
\begin{equation}
\boldsymbol{x}_{t-1}
=
\tilde{\boldsymbol{\mu}}_t(\boldsymbol{x}_t,\hat{\boldsymbol{x}}_{0|t,\boldsymbol{y}})
+
\tilde{\sigma}_t\boldsymbol{z},
\quad
\boldsymbol{z}\sim\mathcal{N}(\mathbf{0},\mathbf{I}),
\label{eq:posterior_reverse_update}
\end{equation}
which has exactly the same form as the standard DDPM reverse update, except that the unconditional clean-signal estimate is replaced by the measurement-conditioned estimate $\hat{\boldsymbol{x}}_{0|t,\boldsymbol{y}}$. Consequently, the remaining task is to derive a tractable expression for $\hat{\boldsymbol{x}}_{0|t,\boldsymbol{y}}$ under the measurement models considered in this paper.


Under the linear measurement model in \eqref{eq:linear-obs-model}, the posterior mean admits the closed-form expression
\begin{equation}
\hat{\boldsymbol{x}}_{0|t,\boldsymbol{y}}
=
\hat{\boldsymbol{x}}_{0|t}
+
\frac{\gamma_t^2}{\gamma_t^2+\sigma_y^2}\,
\mathbf{H}\left(\boldsymbol{y}-\mathbf{H}\hat{\boldsymbol{x}}_{0|t}\right),
\label{eq:linear_posterior_mean}
\end{equation}
where $\hat{\boldsymbol{x}}_{0|t}$ is the unconditional DDPM estimate from \eqref{eq:x0_estimate_from_noise}, and $\gamma_t=\sqrt{\frac{1-\bar{\alpha}_t}{\bar{\alpha}_t}}$. The derivation is given in Appendix~\ref{app:linear_posterior_mean}.



We next consider the quantized measurement model in \eqref{eq:quant_model}. For each measured coordinate $i\in\Omega$, let the quantized measurement $y_i$ correspond to the interval $(l_i,u_i]$. The posterior mean admits the closed-form expression
\begin{equation}
\hat{\boldsymbol{x}}_{0|t,\boldsymbol{y}}
=
\hat{\boldsymbol{x}}_{0|t}
+
\frac{\gamma_t^2}{s_t}\mathbf{H}\boldsymbol{\Delta},
\label{eq:quant_posterior_mean}
\end{equation}
where
\begin{equation}
s_t=\sqrt{\gamma_t^2+\sigma_y^2},
\label{eq:quant_st}
\end{equation}
and $\boldsymbol{\Delta}\in\mathbb{R}^{N}$ is defined element-wise by
\begin{equation}
\Delta_i=
\begin{cases}
\dfrac{\phi(a_i)-\phi(b_i)}{\Phi(b_i)-\Phi(a_i)}, & i\in\Omega,\\[1.2ex]
0, & i\notin\Omega,
\end{cases}
\label{eq:quant_delta}
\end{equation}
where
\begin{equation}
a_i=\frac{l_i-(\hat{\boldsymbol{x}}_{0|t})_i}{s_t},
\quad
b_i=\frac{u_i-(\hat{\boldsymbol{x}}_{0|t})_i}{s_t}.
\label{eq:quant_ab}
\end{equation}
Here, $\phi(\cdot)$ and $\Phi(\cdot)$ denote the PDF and CDF of the standard normal distribution, respectively. The derivation is given in Appendix~\ref{app:quant_posterior_mean}.


Once $\hat{\boldsymbol{x}}_{0|t,\boldsymbol{y}}$ is obtained from either \eqref{eq:linear_posterior_mean} or \eqref{eq:quant_posterior_mean}, each reverse step is completed by evaluating the corrected mean \eqref{eq:posterior_reverse_mean_def} and then sampling according to \eqref{eq:posterior_reverse_update}. The overall GSC reconstruction procedure is summarized in Algorithm~\ref{alg:unified_spectrum}.

\begin{algorithm}[!t]
\caption{Unified Generative Spectrum Cartography (GSC)}
\label{alg:unified_spectrum}
\small
\begin{algorithmic}[1]
\STATE \textbf{Input:} measurement $\boldsymbol{y}$, mask $\mathbf{H}$, pretrained diffusion model $\boldsymbol{\epsilon}_\theta$, diffusion schedule, variance parameters $\{\gamma_t^2\}_{t=1}^{T}$, noise variance $\sigma_y^2$, and quantization parameters (if applicable)
\STATE \textbf{Initialize:} sample $\boldsymbol{x}_T\sim\mathcal{N}(\mathbf{0},\mathbf{I})$
\FOR{$t=T,T-1,\ldots,1$}
    \STATE Compute the unconditional estimate $\hat{\boldsymbol{x}}_{0|t}$ via \eqref{eq:x0_estimate_from_noise}
    \IF{linear measurements}
        \STATE Compute $\hat{\boldsymbol{x}}_{0|t,\boldsymbol{y}}$ via \eqref{eq:linear_posterior_mean}
    \ELSE
        \STATE Compute $s_t$, $a_i$, $b_i$, and $\Delta_i$ via \eqref{eq:quant_st}--\eqref{eq:quant_delta}
        \STATE Compute $\hat{\boldsymbol{x}}_{0|t,\boldsymbol{y}}$ via \eqref{eq:quant_posterior_mean}
    \ENDIF
    \STATE Compute $\tilde{\boldsymbol{\mu}}_t(\boldsymbol{x}_t,\hat{\boldsymbol{x}}_{0|t,\boldsymbol{y}})$ via \eqref{eq:posterior_reverse_mean_def}
    \STATE Sample $\boldsymbol{x}_{t-1}$ via \eqref{eq:posterior_reverse_update}
\ENDFOR
\STATE \textbf{Output:} reconstructed spectrum map $\hat{\boldsymbol{x}}_0$
\end{algorithmic}
\end{algorithm}

\section{Uncertainty-Aware Active Sensing}
\label{sec:uncertainty sensing}

This section develops an uncertainty-aware active sensing strategy within the proposed GSC framework. Specifically, we first estimate an empirical uncertainty map from an ensemble of stochastic reconstructions generated by GSC, and then select a small number of additional sensing locations that are both informative and spatially well distributed. In this way, reconstruction and sensing are integrated within a tractable closed-loop framework.

\subsection{Posterior Uncertainty Estimation}

Since the diffusion model represents a distribution rather than a single point estimate, uncertainty can be quantified by repeated posterior-guided reconstruction under fixed measurements. Specifically, given the current measurement vector $\boldsymbol{y}$ and masking matrix $\mathbf{H}$, we run the GSC reconstruction procedure in Algorithm~1 independently $M$ times with different random seeds for the initial noise and stochastic reverse transitions. This yields an ensemble of reconstructed spectrum maps,
\begin{equation}
\hat{\mathcal{X}}=
\left\{\hat{\mathbf{X}}^{(m)}\right\}_{m=1}^{M},
\quad
\hat{\mathbf{X}}^{(m)}\in\mathbb{R}^{I\times J},
\label{eq:posterior_ensemble}
\end{equation}
where $\hat{\mathbf{X}}^{(m)}$ denotes the $m$-th reconstructed PSD map obtained under the same measurements. Here, the subscripts $i$ and $j$ index the spatial grid location in the reconstructed map.

\begin{figure*}[!t]
    \centering
    \captionsetup[subfigure]{font=scriptsize}
    \subfloat[]{
        \includegraphics[width=0.3\textwidth]{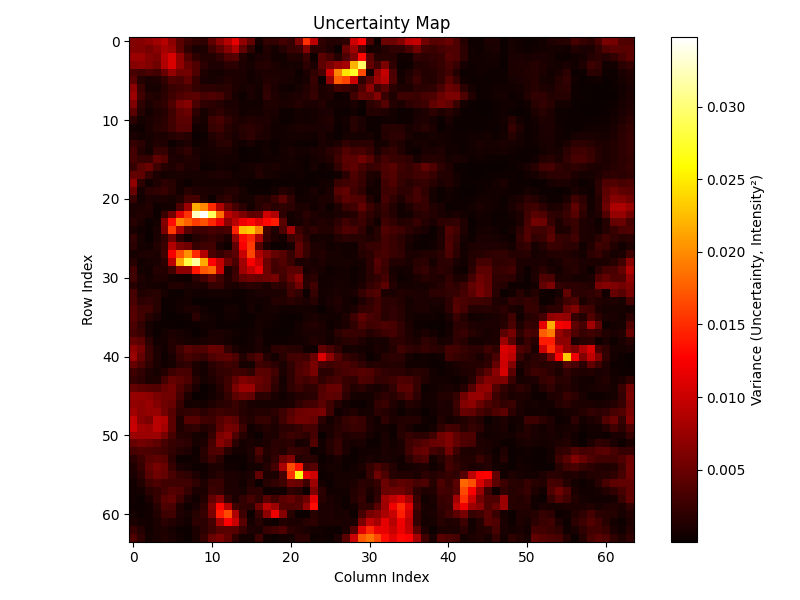}
        \label{fig:uncertainty_map_fig}
    }
    \hfill
    \subfloat[]{
        \includegraphics[width=0.3\textwidth]{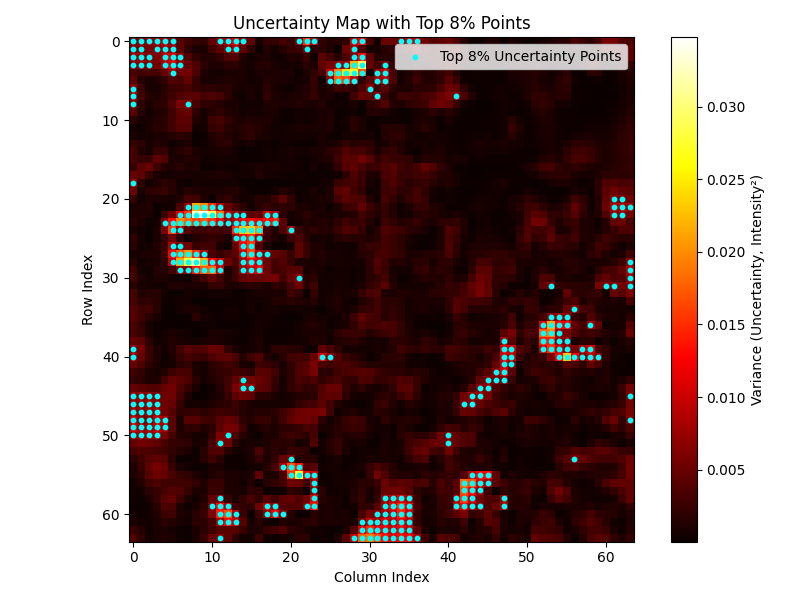}
        \label{fig:top_uncertainty_points}
    }
    \hfill
    \subfloat[]{
        \includegraphics[width=0.3\textwidth]{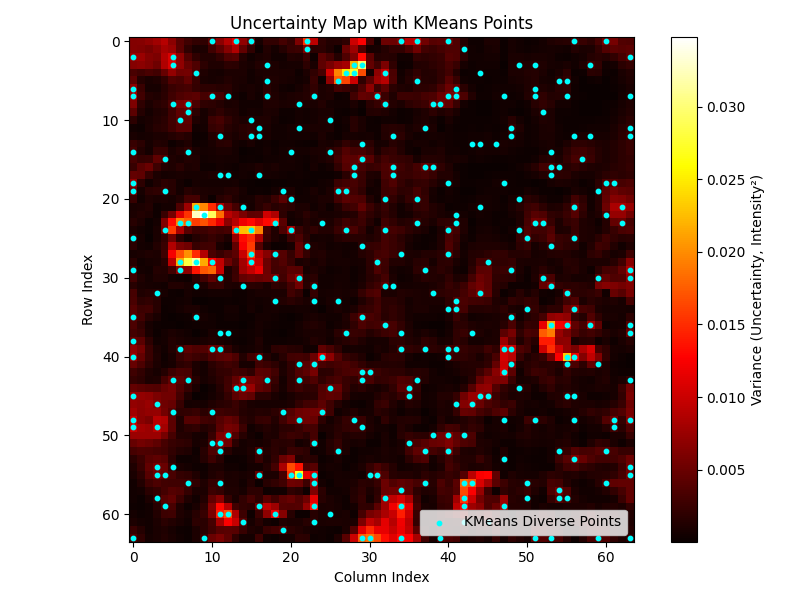}
        \label{fig:top_uncertainty_points_kmean}
    }
    \caption{\small
        Illustration of the proposed uncertainty-aware active sensing strategy.
        (a) Empirical uncertainty map $V_{ij}$ estimated from repeated measurement-conditioned reconstructions.
        (b) Highest-uncertainty candidate locations, showing that uncertainty hotspots tend to cluster spatially.
        (c) New sensing locations selected by the proposed diversity-aware selection strategy based on K-means clustering, which improves spatial coverage across uncertain regions.
    }
    \label{fig:uncertainty_three}
\end{figure*}

From this ensemble, we compute the empirical posterior mean map and uncertainty map, whose entries are
\begin{equation}
\bar{\mathbf{X}}_{ij}
=
\frac{1}{M}\sum_{m=1}^{M}\hat{\mathbf{X}}^{(m)}_{ij},
\label{eq:posterior_mean_empirical}
\end{equation}
\begin{equation}
V_{ij}
=
\frac{1}{M-1}\sum_{m=1}^{M}
\left(
\hat{\mathbf{X}}^{(m)}_{ij}-\bar{\mathbf{X}}_{ij}
\right)^2.
\label{eq:uncertainty_map}
\end{equation}
Large values of $V_{ij}$ indicate greater uncertainty in the reconstruction at location $(i,j)$. Therefore, the uncertainty map directly highlights regions that remain weakly constrained by the available measurements. As illustrated in Fig.~\ref{fig:uncertainty_three}(a), it provides a spatial summary of the uncertainty distribution under the current measurement condition.

\subsection{Diversity-Aware Sensing-Location Selection}

Under a limited sensing budget, uncertainty provides a natural indicator of the sensing value of a location. Intuitively, if a location has high uncertainty, then the current reconstruction remains ambiguous there, and acquiring an additional measurement at that location is more likely to improve subsequent reconstruction quality.

However, directly selecting the top-$Q$ most uncertain locations is often inefficient. As shown in Fig.~\ref{fig:uncertainty_three}(b), high-uncertainty entries tend to cluster spatially, for example around poorly observed source boundaries or localized interference regions. Sampling many nearby locations in such a cluster may provide redundant information and thus waste the sensing budget. For this reason, uncertainty alone is not sufficient; spatial diversity must also be taken into account.

Let $h_{ij}\in\{0,1\}$ denote the entry of the grid-form sampling mask associated with $\mathbf{H}$, where $h_{ij}=0$ indicates that the grid entry indexed by $i$ and $j$ is currently unmeasured. Since new measurements can only be acquired at unmeasured locations, the active sensing task is to select $Q$ additional sensing locations that are both informative and spatially well distributed.

To improve spatial coverage, we adopt a diversity-aware selection strategy based on K-means clustering \cite{macqueen1967kmeans}. For each unmeasured grid entry, we construct a three-dimensional feature vector consisting of the normalized spatial coordinates and the normalized uncertainty. Specifically, let $V_{\max}$ denote the maximum uncertainty over all currently unmeasured grid entries, and define the normalized uncertainty as $\tilde{v}_{ij}=V_{ij}/V_{\max}$. We then define
\begin{equation}
\mathbf{d}_{ij}
=
\left[
\frac{i-1}{I-1},\;
\frac{j-1}{J-1},\;
\tilde{v}_{ij}
\right],
\quad h_{ij}=0.
\label{eq:active_feature}
\end{equation}
Collecting all such feature vectors gives the candidate set
\begin{equation}
\mathcal{D}
=
\left\{
\mathbf{d}_{ij}: h_{ij}=0
\right\}.
\label{eq:candidate_set}
\end{equation}
We then partition $\mathcal{D}$ into $Q$ clusters by K-means clustering. For each cluster $\mathcal{C}_k$, we select the candidate with the largest uncertainty and denote the corresponding sensing location by $\mathbf{p}_k$. The resulting set of selected sensing locations is
\begin{equation}
\mathcal{P}_{\mathrm{new}}
=
\left\{
\mathbf{p}_1,\ldots,\mathbf{p}_Q
\right\}.
\label{eq:new_sensing_set}
\end{equation}
Compared with naive top-$Q$ selection, this procedure preserves the motivation of uncertainty-aware sensing while preventing the selected locations from collapsing into a single hotspot. As illustrated in Fig.~\ref{fig:uncertainty_three}(b)--(c), it encourages the sensing budget to cover multiple uncertain regions.
\subsection{Active Sensing Algorithm}

After $\mathcal{P}_{\mathrm{new}}$ is determined, new measurements are acquired at these locations and incorporated into the measurement vector and masking matrix. The updated measurement vector and masking matrix are then fed back to the reconstruction stage, thereby forming a closed-loop reconstruction-and-sensing process. The overall procedure is summarized in Algorithm~\ref{alg:uncertainty_guided}.

\begin{algorithm}[!t]
\caption{Uncertainty-Aware Active Sensing}
\label{alg:uncertainty_guided}
\small
\begin{algorithmic}[1]
\STATE \textbf{Input:} current measurement vector $\boldsymbol{y}$, current masking matrix $\mathbf{H}$, ensemble size $M$, sensing budget $Q$, and all reconstruction parameters required by Algorithm~\ref{alg:unified_spectrum}
\STATE \textbf{Posterior Sampling:} Run Algorithm~\ref{alg:unified_spectrum} independently $M$ times to obtain the ensemble $\mathcal{X}$
\STATE \textbf{Uncertainty Estimation:} Compute the empirical posterior mean and uncertainty map via \eqref{eq:posterior_mean_empirical} and \eqref{eq:uncertainty_map}
\STATE \textbf{Candidate Construction:} For each unmeasured grid entry, compute the normalized uncertainty and form the feature set $\mathcal{D}$
\STATE \textbf{Clustering:} Apply K-means to partition $\mathcal{D}$ into $Q$ clusters $\{\mathcal{C}_1,\ldots,\mathcal{C}_Q\}$
\STATE \textbf{Selection:} Select one sensing location from each cluster according to the largest uncertainty, and obtain $\mathcal{P}_{\mathrm{new}}$
\STATE \textbf{Measurement Update:} Acquire new measurements at $\mathcal{P}_{\mathrm{new}}$ and update the measurement vector and masking matrix to $\boldsymbol{y}^{+}$ and $\mathbf{H}^{+}$
\STATE \textbf{Output:} selected sensing locations $\mathcal{P}_{\mathrm{new}}$, updated measurement vector $\boldsymbol{y}^{+}$, and updated masking matrix $\mathbf{H}^{+}$
\end{algorithmic}
\end{algorithm}

\section{Experiments}
\label{sec:experiment}

This section evaluates the proposed Generative Spectrum Cartography (GSC) framework on two fully simulated testbeds: a simulated electromagnetic map dataset and a high-fidelity simulated satellite monitoring scenario. The experiments are designed to assess two aspects of the proposed framework: reconstruction performance under sparse, noisy, and low-bit quantized measurements, and sensing efficiency of the uncertainty-aware active sensing strategy.

\textbf{Baselines.}
We use different baseline sets according to the measurement regime.

\paragraph{Linear-measurement setting.}
For the linear setting, we compare GSC with four representative baselines:
(1) Inverse Distance Weighting (IDW) \cite{denkovski2012reliability}, a classical interpolation method;
(2) Nasdac \cite{shrestha2022deep}, a neural-network-based spectrum cartography model;
(3) Dowjons \cite{shrestha2022deep}, another learned spectrum cartography approach with spatially informed loss design; and
(4) Diffusion Posterior Sampling (DPS) \cite{chung2023diffusion}, a representative diffusion-based inverse solver that incorporates measurements through likelihood-gradient guidance.

\paragraph{Quantized-measurement setting.}
For low-bit quantized measurements (1-bit, 2-bit, and 3-bit), we compare against the UNN network \cite{11127065}, a training-free spectrum cartography method that exploits the inductive bias of untrained neural architectures. We use UNN as a representative baseline for heavily quantized spectrum reconstruction.

\paragraph{Active sensing setting.}
For active sensing, we compare the proposed uncertainty-aware sensing strategy with random sensing under matched additional sensing budgets.

\textbf{Evaluation metrics.}
We report peak signal-to-noise ratio (PSNR) and Learned Perceptual Image Patch Similarity (LPIPS) \cite{zhang2018unreasonable}. PSNR is treated as the primary fidelity metric, while LPIPS is included as a complementary structural/perceptual indicator of reconstruction quality. Higher PSNR and lower LPIPS indicate better performance.

\subsection{Simulated Electromagnetic Map Dataset}
\label{subsec:simulated}

\textbf{Dataset construction.}
Following the hybrid electromagnetic map simulator in \cite{shrestha2022deep}, the simulated electromagnetic map dataset is generated by combining a Spatial Loss Function (SLF) component and a Power Spectral Density (PSD) component. The SLF component captures distance-dependent path loss and log-normal shadowing, where the shadowing term follows a zero-mean Gaussian process with spatial correlation controlled by a decorrelation distance. The PSD component is modeled as a sum of randomly scaled sinc functions, whose center frequencies, sidelobe widths, and power amplitudes are sampled from predefined uniform distributions. The resulting electromagnetic field over a \(50 \times 50\) support is given by
\begin{equation}
\mathbf{X}_0 = \sum_{r=1}^{R} \mathbf{S}_r \odot \boldsymbol{c_r} + \boldsymbol{n},
\label{eq:em_map_model}
\end{equation}
where $\mathbf{S}_r$ and $\boldsymbol{c_r}$ denote the SLF and PSD components of the \(r\)-th source, respectively, $\boldsymbol{n}$ denotes additive noise.

\begin{figure}[htbp]
    \centering
    \includegraphics[width=0.4\linewidth]{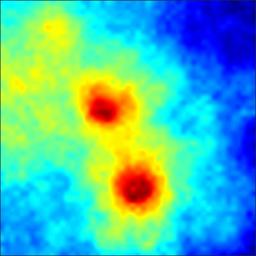}
    \caption{Representative fixed-frequency slice from the simulated electromagnetic map dataset. The map exhibits heterogeneous interference structure and smooth background variation, providing a nontrivial test case for spectrum cartography under sparse measurements.}
    \label{fig:em_map_slice}
\end{figure}

The dataset contains 100{,}000 simulated electromagnetic maps. Transmitter locations are uniformly distributed over the spatial domain. To model multi-source interference, we superimpose independently generated SLF components, with the number of sources ranging from $R\in\{1,2,\ldots,10\}$. Each simulated environment yields a multi-frequency electromagnetic tensor, from which fixed-frequency 2D slices are extracted as target PSD maps in the present work. These 100{,}000 samples are used to train the diffusion prior and capture the statistical structure of electromagnetic field distributions. A representative 2D slice is shown in Fig.~\ref{fig:em_map_slice}.

\begin{figure}[htbp]
    \centering
    \includegraphics[width=1\linewidth, trim=3 3 3 3, clip]{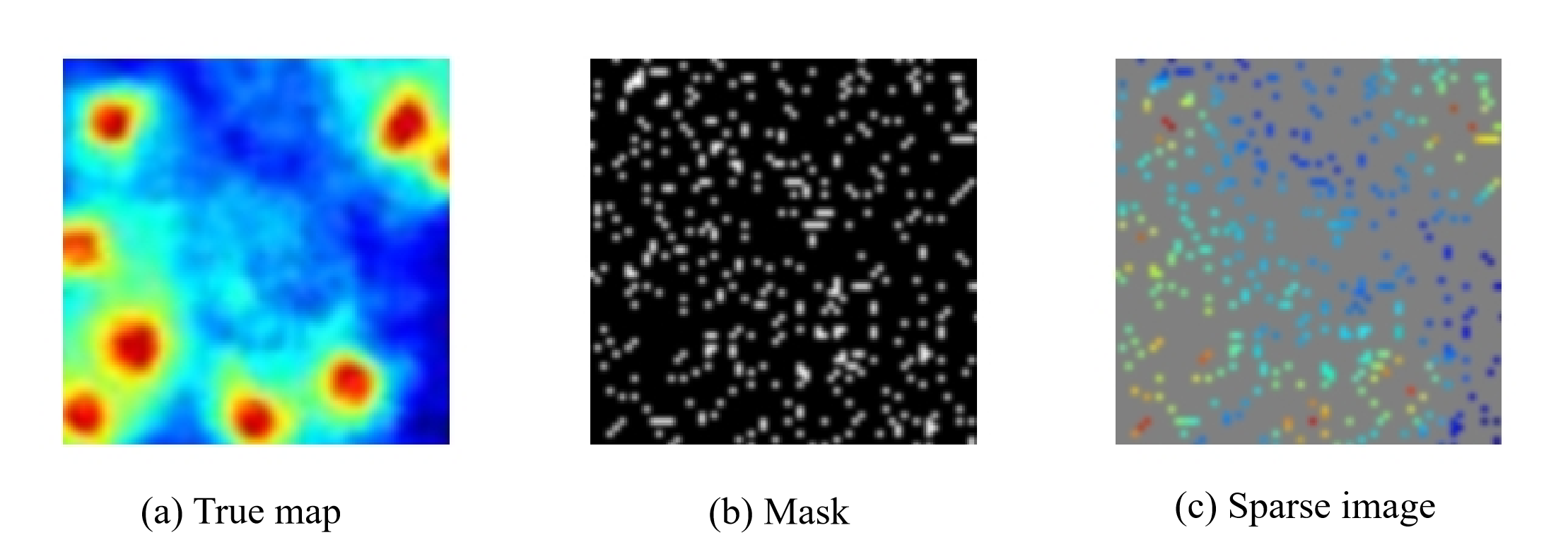}
    \caption{Example of the sparse-measurement protocol. From left to right: ground-truth map, binary sampling mask, and sparse measurement obtained after masking. The experiments use random masks with measurement ratios \(\rho=0.20, 0.15, 0.10,\) and \(0.05\).}
    \label{fig:mask}
\end{figure}

\begin{figure}[htbp]
    \centering
    \includegraphics[width=1\linewidth, trim=10 3 8 3, clip]{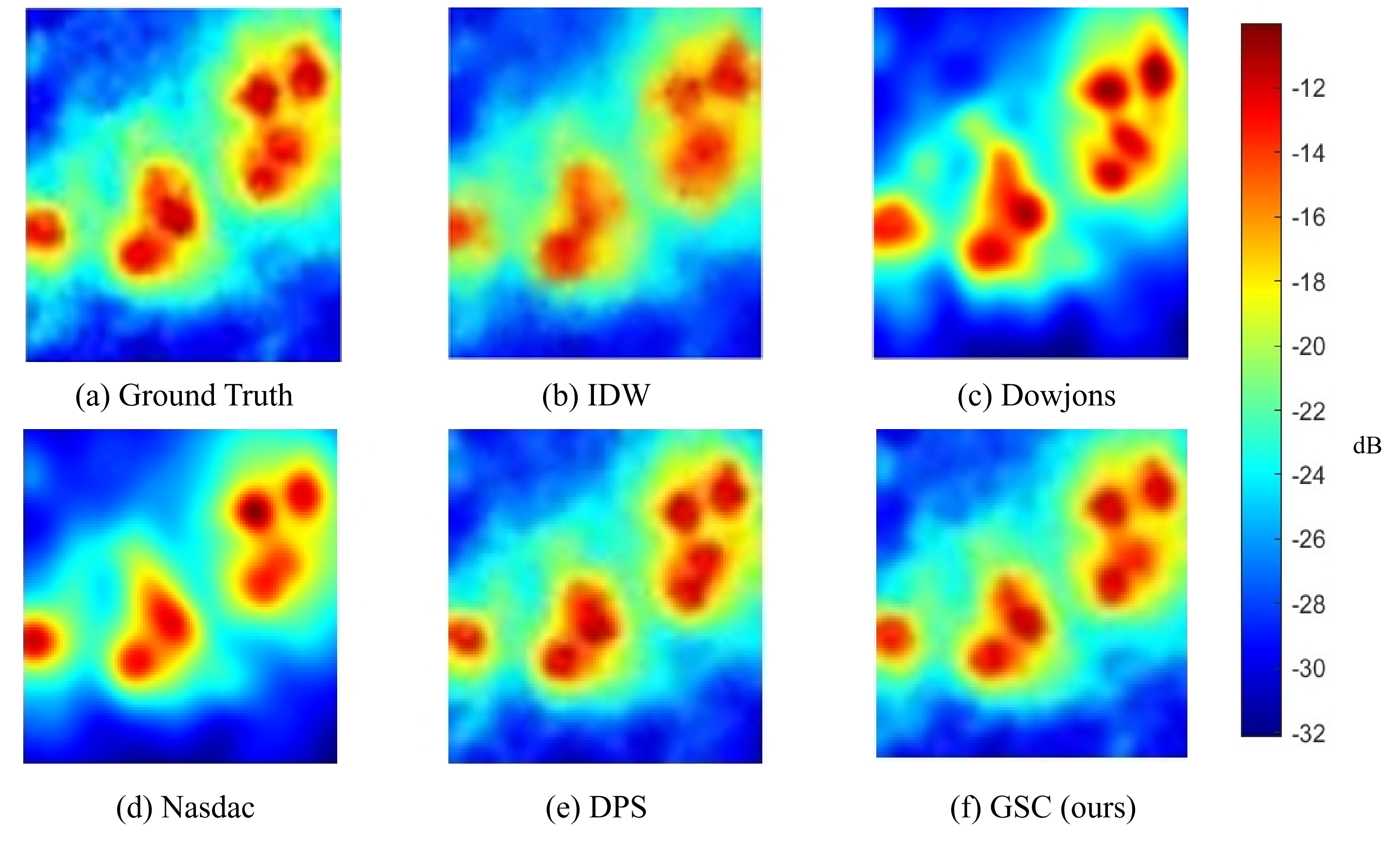}
    \caption{Representative reconstruction results under measurement ratio \(\rho=0.15\) in the noise-free setting. From left to right and top to bottom, the panels show the ground truth, IDW, Dowjons, Nasdac, DPS, and the proposed GSC. GSC better preserves localized spatial structure and sharp transitions than the competing methods.}
    \label{fig:85mask}
\end{figure}

\begin{figure}[htbp]
    \centering
    \includegraphics[width=1\linewidth, trim=10 3 8 3, clip]{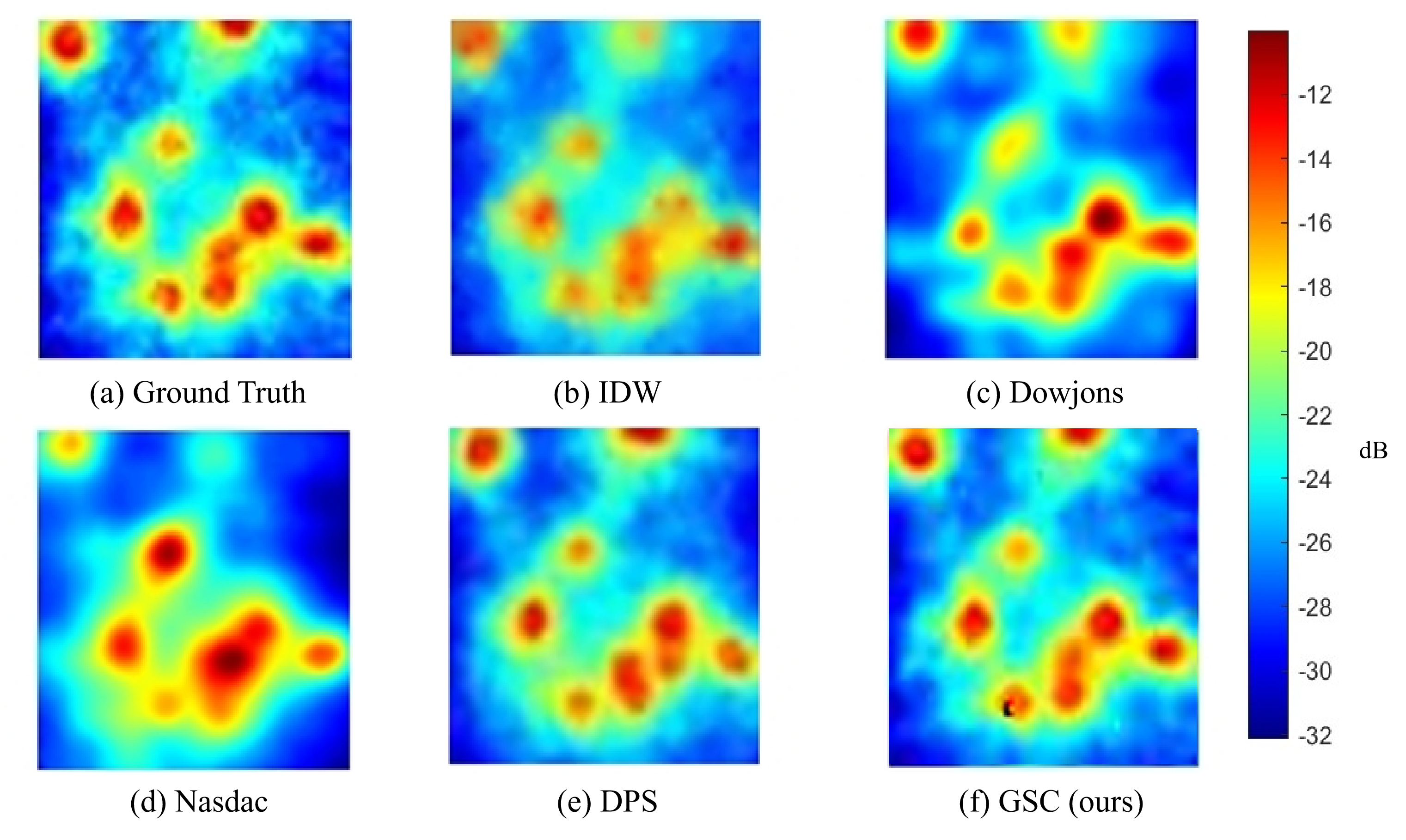}
    \caption{Representative reconstruction results under measurement ratio \(\rho=0.05\) in the noise-free setting. From left to right and top to bottom, the panels show the ground truth, IDW, Dowjons, Nasdac, DPS, and the proposed GSC. Even in this highly sparse regime, GSC better recovers major spatial structures and interference patterns.}
    \label{fig:95mask}
\end{figure}

\textbf{Experimental setup.}
After training, we generate an additional test set consisting of 100 simulated electromagnetic maps for evaluation. Sparse measurements are produced by applying random masks with measurement ratios \(\rho \in \{0.20,0.15,0.10,0.05\}\). All quantitative results reported in this subsection are averaged over these 100 test maps. A representative sparse-measurement example is shown in Fig.~\ref{fig:mask}. Unless otherwise specified, reconstruction is first evaluated in the noise-free setting, and representative visual results for \(\rho=0.15\) and \(\rho=0.05\) are shown in Figs.~\ref{fig:85mask} and~\ref{fig:95mask}, respectively.

\textbf{Noise-free reconstruction results.}
Figures~\ref{fig:85mask} and~\ref{fig:95mask} show that classical interpolation methods such as IDW tend to over-smooth the reconstructed maps, especially at small measurement ratios, leading to the loss of local structure and interference details. The learned cartography baselines provide moderate improvements, but their reconstructions remain noticeably less faithful than those of the generative approaches in the sparse regime. By contrast, GSC better preserves localized interference patterns and sharp spatial transitions by combining a learned diffusion prior with measurement-conditioned posterior correction.

The quantitative results in Table~\ref{tab:noisefree_combined} confirm this trend. As the measurement ratio decreases from \(\rho=0.20\) to \(\rho=0.05\), all methods exhibit lower PSNR and higher LPIPS, reflecting the increased difficulty of the reconstruction task. Among the interpolation and deterministic learning baselines, performance remains limited, with PSNR mostly in the 16--18 dB range and LPIPS around 0.45--0.51. DPS substantially improves performance by introducing a generative prior. GSC consistently outperforms DPS across all measurement ratios: for example, at \(\rho=0.20\), GSC reaches 37.22 dB PSNR and 0.010 LPIPS, compared with 34.47 dB and 0.018 for DPS. Even at the most challenging setting \(\rho=0.05\), GSC still attains 27.51 dB PSNR and 0.058 LPIPS. These results support the effectiveness of the proposed closed-form posterior mean update over likelihood-gradient-based correction in sparse reconstruction.

\begin{table*}[t]
\centering
\caption{Average noise-free reconstruction performance on 100 held-out simulated electromagnetic maps. Higher PSNR and lower LPIPS indicate better performance. Best results are shown in bold.}
\label{tab:noisefree_combined}
\setlength{\tabcolsep}{5pt}
\begin{tabular}{
l
S[table-format=2.2] S[table-format=2.2] S[table-format=2.2] S[table-format=2.2]
S[table-format=1.3] S[table-format=1.3] S[table-format=1.3] S[table-format=1.3]
}
\toprule
\multirow{2}{*}{Method} &
\multicolumn{4}{c}{PSNR (dB) $\uparrow$} &
\multicolumn{4}{c}{LPIPS $\downarrow$} \\
\cmidrule(lr){2-5} \cmidrule(lr){6-9}
& {$\rho=0.20$} & {$\rho=0.15$} & {$\rho=0.10$} & {$\rho=0.05$}
& {$\rho=0.20$} & {$\rho=0.15$} & {$\rho=0.10$} & {$\rho=0.05$} \\
\midrule
IDW     & 18.49 & 17.99 & 16.62 & 16.21 & 0.460 & 0.470 & 0.510 & 0.500 \\
Nasdac  & 17.00 & 16.58 & 15.52 & 16.50 & 0.460 & 0.460 & 0.480 & 0.460 \\
Dowjons & 16.52 & 16.16 & 16.39 & 17.00 & 0.450 & 0.470 & 0.460 & 0.400 \\
DPS     & 34.47 & 32.60 & 29.87 & 25.26 & 0.018 & 0.024 & 0.038 & 0.072 \\
GSC (ours) & {\bfseries 37.22} & {\bfseries 34.60} & {\bfseries 31.69} & {\bfseries 27.51}
           & {\bfseries 0.010} & {\bfseries 0.017} & {\bfseries 0.029} & {\bfseries 0.058} \\
\bottomrule
\end{tabular}
\end{table*}

\textbf{Noisy reconstruction results.}
We next evaluate reconstruction under additive Gaussian noise with variance \(\sigma^2=0.05\). The quantitative results in Table~\ref{tab:noisy_combined} show that GSC maintains a clear advantage over all baselines in the noisy setting. At \(\rho=0.20\), IDW achieves only 9.09 dB PSNR, while DPS improves this to 30.19 dB and GSC further improves it to 34.41 dB. Similar trends are observed at \(\rho=0.15\) and \(\rho=0.10\), where GSC consistently attains the highest PSNR. The same pattern holds for LPIPS: GSC yields lower structural/perceptual error than DPS across all measurement ratios, for example 0.024 versus 0.037 at \(\rho=0.20\) and 0.027 versus 0.045 at \(\rho=0.15\).

To further examine noise sensitivity, we fix the measurement ratio at \(\rho=0.15\) and vary the noise variance over \(\sigma^2 \in \{0.05,0.1,0.2,0.5\}\). As reported in Table~\ref{tab:noise_level_comparison}, PSNR decreases from 33.58 dB to 23.53 dB and LPIPS increases from 0.027 to 0.151 as the noise level increases. Although performance degrades under heavier noise, the degradation is gradual rather than abrupt, indicating that the learned generative prior provides stable denoising support across a broad range of noise levels.

\begin{table*}[t]
\centering
\caption{Average reconstruction performance on 100 held-out simulated electromagnetic maps under additive Gaussian noise (\(\sigma^2=0.05\)). Higher PSNR and lower LPIPS indicate better performance. Best results are shown in bold.}
\label{tab:noisy_combined}
\setlength{\tabcolsep}{5pt}
\begin{tabular}{
l
S[table-format=2.2] S[table-format=2.2] S[table-format=2.2] S[table-format=2.2]
S[table-format=1.3] S[table-format=1.3] S[table-format=1.3] S[table-format=1.3]
}
\toprule
\multirow{2}{*}{Method} &
\multicolumn{4}{c}{PSNR (dB) $\uparrow$} &
\multicolumn{4}{c}{LPIPS $\downarrow$} \\
\cmidrule(lr){2-5} \cmidrule(lr){6-9}
& {$\rho=0.20$} & {$\rho=0.15$} & {$\rho=0.10$} & {$\rho=0.05$}
& {$\rho=0.20$} & {$\rho=0.15$} & {$\rho=0.10$} & {$\rho=0.05$} \\
\midrule
IDW     &  9.09 &  8.83 &  8.58 &  8.34 & 0.693 & 0.727 & 0.766 & 0.812 \\
Nasdac  & 19.04 & 19.08 & 18.90 & 17.24 & 0.223 & 0.215 & 0.220 & 0.244 \\
Dowjons & 17.97 & 18.85 & 18.62 & 17.38 & 0.231 & 0.212 & 0.215 & 0.235 \\
DPS     & 30.19 & 28.80 & 26.66 & 23.10 & 0.037 & 0.045 & 0.060 & 0.097 \\
GSC (ours) & {\bfseries 34.41} & {\bfseries 33.58} & {\bfseries 32.31} & {\bfseries 28.65}
           & {\bfseries 0.024} & {\bfseries 0.027} & {\bfseries 0.032} & {\bfseries 0.059} \\
\bottomrule
\end{tabular}
\end{table*}

\begin{table}[t]
\centering
\caption{Sensitivity of GSC to different Gaussian noise variances at measurement ratio \(\rho=0.15\) on the simulated electromagnetic map dataset. Higher PSNR and lower LPIPS indicate better performance.}
\label{tab:noise_level_comparison}
\setlength{\tabcolsep}{6pt}
\begin{tabular}{
S[table-format=1.2]
S[table-format=2.2]
S[table-format=1.3]
}
\toprule
{$\sigma^2$} & {PSNR (dB) $\uparrow$} & {LPIPS $\downarrow$} \\
\midrule
0.05 & 33.58 & 0.027 \\
0.10 & 30.36 & 0.052 \\
0.20 & 27.03 & 0.096 \\
0.50 & 23.53 & 0.151 \\
\bottomrule
\end{tabular}
\end{table}

\textbf{Quantized reconstruction results.}
We further evaluate GSC under 1-bit, 2-bit, and 3-bit scalar quantization using the same sparse-measurement protocol as in the linear setting. Table~\ref{tab:results_quantized} reports the quantitative comparison with UNN. The results show that low-bit spectrum reconstruction remains highly challenging for the baseline: UNN yields PSNR values around 10--11 dB and LPIPS values above 0.46 across all quantization levels.

In contrast, GSC achieves substantially better performance at every bit-depth. At \(\rho=0.15\), GSC attains 20.98 dB PSNR and 0.183 LPIPS under 1-bit quantization, compared with 10.95 dB and 0.491 for UNN. As the quantization resolution increases, the reconstruction quality of GSC further improves: at \(\rho=0.15\), it reaches 25.20 dB and 26.72 dB PSNR for 2-bit and 3-bit quantization, respectively. Similar gains are observed at \(\rho=0.20\). These results are consistent with the advantage of explicitly incorporating the quantization likelihood through Gaussian interval probabilities, rather than relying on gradient-based corrections that are difficult to apply in low-bit nonlinear sensing regimes.

\begin{table*}[t]
\centering
\caption{Average reconstruction performance on 100 held-out simulated electromagnetic maps under low-bit quantization. Higher PSNR and lower LPIPS indicate better performance. Best results are shown in bold. Gray rows correspond to the proposed method.}
\label{tab:results_quantized}
\setlength{\tabcolsep}{6pt}
\renewcommand{\arraystretch}{1.08}
\begin{tabular}{
ll
S[table-format=2.2]
S[table-format=1.3]
S[table-format=2.2]
S[table-format=1.3]
}
\toprule
\multirow{2}{*}{Bit-depth} & \multirow{2}{*}{Method}
& \multicolumn{2}{c}{$\rho=0.20$}
& \multicolumn{2}{c}{$\rho=0.15$} \\
\cmidrule(lr){3-4} \cmidrule(lr){5-6}
& & {PSNR (dB) $\uparrow$} & {LPIPS $\downarrow$}
  & {PSNR (dB) $\uparrow$} & {LPIPS $\downarrow$} \\
\midrule

\multirow{2}{*}{1-bit}
& UNN & 11.27 & 0.476 & 10.95 & 0.491 \\
\rowcolor{black!6}
& GSC (ours) & {\bfseries 21.67} & {\bfseries 0.170} & {\bfseries 20.98} & {\bfseries 0.183} \\
\addlinespace[3pt]

\multirow{2}{*}{2-bit}
& UNN & 11.00 & 0.489 & 10.84 & 0.505 \\
\rowcolor{black!6}
& GSC (ours) & {\bfseries 25.14} & {\bfseries 0.098} & {\bfseries 25.20} & {\bfseries 0.098} \\
\addlinespace[3pt]

\multirow{2}{*}{3-bit}
& UNN & 11.26 & 0.469 & 10.72 & 0.494 \\
\rowcolor{black!6}
& GSC (ours) & {\bfseries 27.75} & {\bfseries 0.067} & {\bfseries 26.72} & {\bfseries 0.086} \\
\bottomrule
\end{tabular}
\end{table*}

\textbf{Active sensing results.}
We next evaluate the closed-loop active sensing strategy introduced in Section~IV. In the linear-measurement setting, we start from an initial measurement ratio $\rho_0=L_0/N=0.10$, where $M_0$ denotes the number of initially measured locations. Based on these initial measurements, we estimate an uncertainty map from repeated measurement-conditioned reconstructions and then select additional sensing locations using either random sensing or the proposed uncertainty-aware strategy. Let $\Delta\rho=Q/N$ denote the additional measurement ratio, where $Q$ is the number of newly selected sensing locations. In the experiments, we consider $\Delta\rho\in\{0.03,0.04,0.05,0.06,0.07\}$. For each additional sensing budget, the corresponding measurements are revealed and merged into the measurement set before reconstruction is rerun on the updated measurements.

\begin{figure}[t]
    \centering
    \includegraphics[width=0.84\columnwidth]{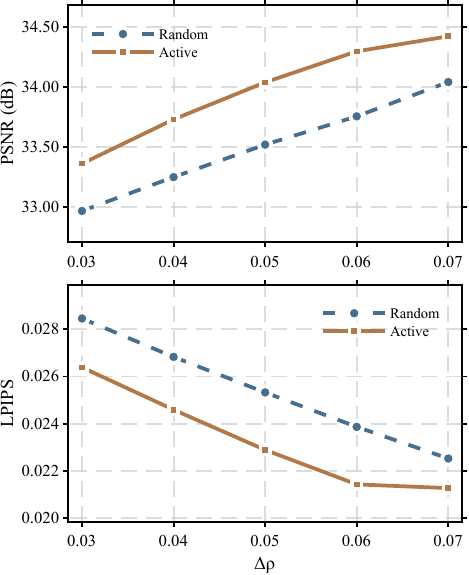}
    \caption{Active sensing performance under linear measurements, starting from an initial measurement ratio \(\rho_0=0.10\). The top subplot shows PSNR and the bottom subplot shows LPIPS as functions of the additional measurement ratio \(\Delta\rho\). The proposed uncertainty-aware strategy consistently outperforms random sensing under matched sensing budgets.}
    \label{fig:active_learning}
\end{figure}

Figure~\ref{fig:active_learning} shows that the proposed strategy consistently outperforms random sensing across all additional sensing budgets, yielding higher PSNR and lower LPIPS. The improvement becomes larger as \(\Delta\rho\) increases, indicating that posterior uncertainty provides useful guidance for identifying measurements that contribute more effectively to reconstruction quality.

\begin{figure*}[t]
    \centering
    \includegraphics[width=0.94\textwidth]{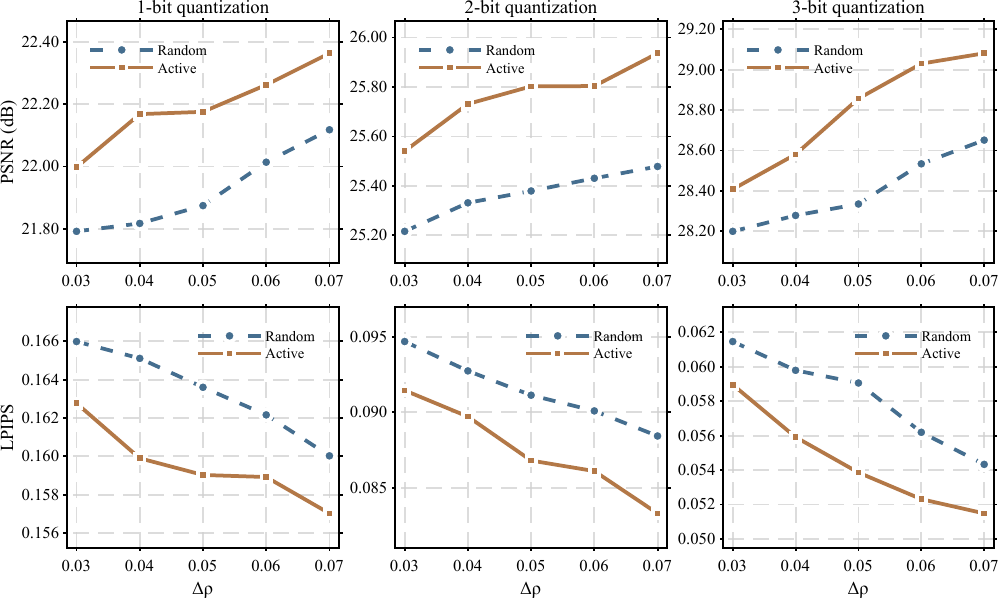}
    \caption{Active sensing performance under low-bit quantized measurements, starting from an initial measurement ratio \(\rho_0=0.20\). The top row shows PSNR and the bottom row shows LPIPS. The three columns correspond to 1-bit, 2-bit, and 3-bit quantization, respectively. Across all bit-depths, the proposed uncertainty-aware strategy consistently improves over random sensing as the additional measurement ratio \(\Delta\rho\) increases.}
    \label{fig:active_learning_quantized}
\end{figure*}

We then consider the quantized setting, where uncertainty estimation is more challenging. In these experiments, the initial measurement ratio is set to \(\rho_0=0.20\), after which additional sensing locations are selected either randomly or by the proposed strategy, their measurements are acquired, and reconstruction is repeated on the augmented measurement set. As shown in Fig.~\ref{fig:active_learning_quantized}, the proposed method maintains a consistent advantage across 1-bit, 2-bit, and 3-bit quantization. Even in the highly challenging 1-bit case, uncertainty-aware sensing still improves over random sensing. As the bit-depth increases, the advantage becomes more pronounced, with consistently higher PSNR and lower LPIPS under all additional sensing budgets.

Overall, these results show that uncertainty guided sensing improves the efficiency of the sensing budget in both linear and quantized regimes. By directing additional measurements toward weakly constrained regions, the proposed strategy yields more accurate and structurally better reconstructions than random sensing.

\subsection{High-Fidelity Simulated Satellite Monitoring Scenario}
\label{subsec:satellite}

\textbf{Satellite simulation setting.}
To further evaluate the proposed reconstruction mechanism under a more realistic and physically detailed simulation setting, we consider electromagnetic maps generated by a high-fidelity simulated satellite monitoring environment in MATLAB. The scenario emulates wide-area monitoring in which a Medium Earth Orbit (MEO) satellite (altitude \(\approx 20{,}000\) km) monitors uplink electromagnetic interference from a Low Earth Orbit (LEO) mega-constellation. The sensing region spans latitudes 27\(^\circ\)N--54\(^\circ\)N and longitudes 96\(^\circ\)E--123\(^\circ\)E, discretized into a \(64\times 64\) support. The LEO constellation operates at an altitude of 550 km with an inclination of 53\(^\circ\), comprising 72 orbital planes with 44 satellites per plane. The simulation includes 300 randomly distributed ground stations communicating with LEO satellites according to a maximum-elevation-angle access rule, with antenna patterns following the ITU-R S.465 standard.

\begin{figure}[htbp]
    \centering
    \includegraphics[width=1\linewidth, trim=3 10 3 10, clip]{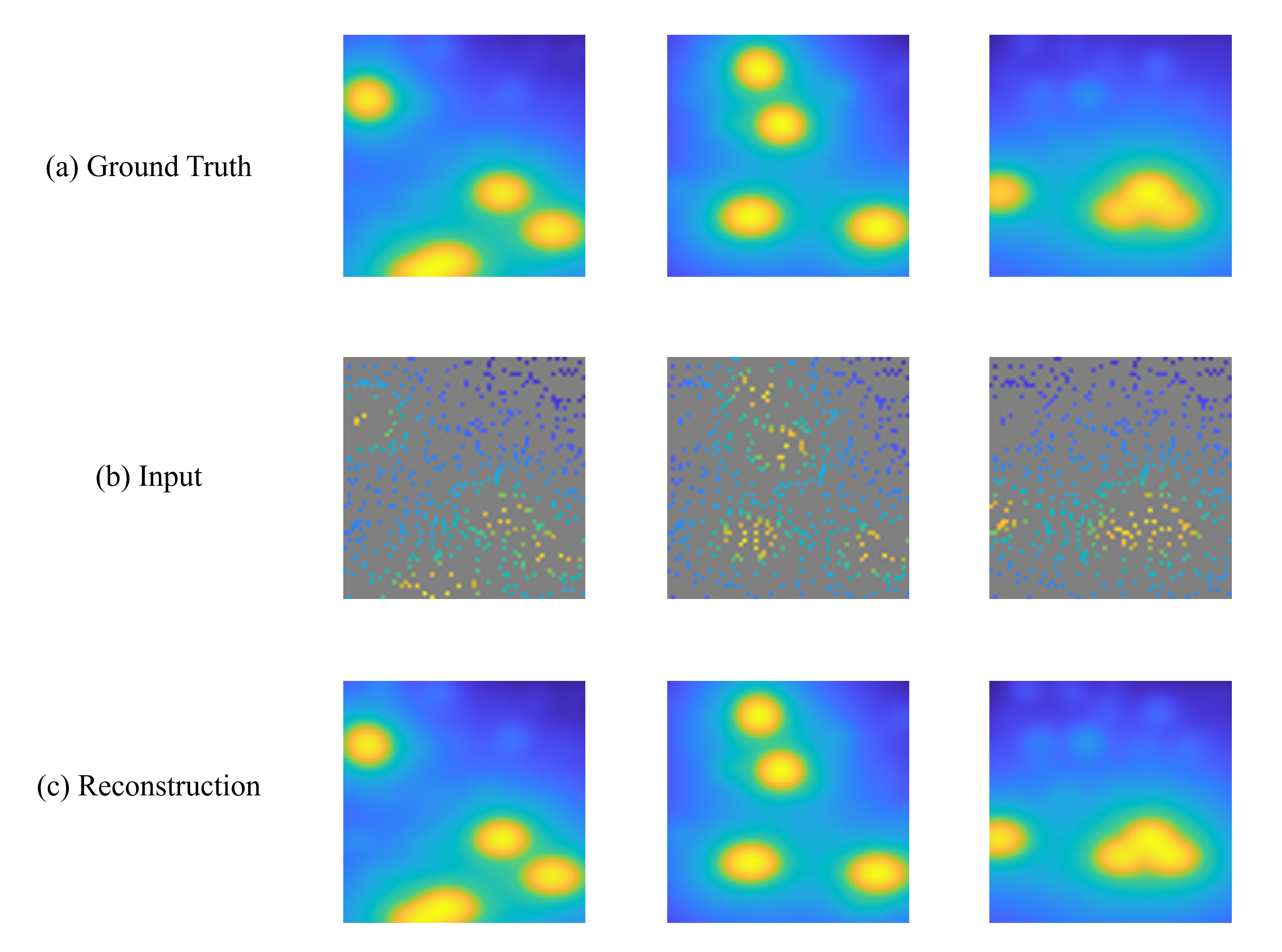}
    \caption{Representative reconstructions in the high-fidelity simulated satellite monitoring scenario under measurement ratio \(\rho=0.10\). Each column corresponds to a different test case, and the three rows show the ground truth, sparse measurement, and GSC reconstruction, respectively.}
    \label{fig:em_map_slice1}
\end{figure}

\textbf{Satellite reconstruction results.}
We evaluate reconstruction under measurement ratios \(\rho=0.20\) and \(\rho=0.10\). A representative result for \(\rho=0.10\) is shown in Fig.~\ref{fig:em_map_slice1}. Even with highly incomplete measurements, GSC recovers the major interference structures and broad spatial variation patterns, yielding reconstructions that remain visually close to the ground truth.

Table~\ref{tab:resultstrue} reports the quantitative comparison with the classical IDW baseline on this simulated satellite dataset. At \(\rho=0.20\), IDW achieves 29.00 dB PSNR and 0.074 LPIPS, whereas GSC reaches 39.07 dB and 0.049 LPIPS. When the measurement ratio is reduced to \(\rho=0.10\), IDW degrades to 25.30 dB PSNR and 0.137 LPIPS, while GSC still maintains 37.35 dB and 0.036 LPIPS. These results provide encouraging evidence that the proposed measurement-conditioned generative reconstruction remains effective in a more realistic simulated satellite monitoring scenario.

\begin{table}[t]
\centering
\caption{Reconstruction performance in the high-fidelity simulated satellite monitoring scenario. Higher PSNR and lower LPIPS indicate better performance. Best results are shown in bold.}
\label{tab:resultstrue}
\setlength{\tabcolsep}{6pt}
\begin{tabular}{
l
S[table-format=2.2]
S[table-format=1.3]
S[table-format=2.2]
S[table-format=1.3]
}
\toprule
\multirow{2}{*}{Method}
& \multicolumn{2}{c}{$\rho=0.20$}
& \multicolumn{2}{c}{$\rho=0.10$} \\
\cmidrule(lr){2-3} \cmidrule(lr){4-5}
& {PSNR (dB) $\uparrow$} & {LPIPS $\downarrow$}
& {PSNR (dB) $\uparrow$} & {LPIPS $\downarrow$} \\
\midrule
IDW & 29.00 & 0.074 & 25.30 & 0.137 \\
GSC (ours) & {\bfseries 39.07} & {\bfseries 0.049} & {\bfseries 37.35} & {\bfseries 0.036} \\
\bottomrule
\end{tabular}
\end{table}

\section{Conclusion}
\label{sec:conclusion}

This paper proposed Generative Spectrum Cartography (GSC), a diffusion-based posterior inference framework for spectrum cartography, to jointly address two coupled challenges: high-fidelity spectrum map reconstruction under sparse, noisy, and low-bit quantized measurements, and efficient sensing under a limited sensing budget. Unlike diffusion-based inverse solvers that rely on likelihood-gradient approximations, GSC directly incorporates measurement information into the reverse diffusion process through closed-form posterior mean updates, yielding a gradient-free and measurement-consistent reconstruction method. These closed-form updates apply to both linear and quantized measurement models. Furthermore, by estimating spatial uncertainty from posterior samples produced by the measurement-conditioned reconstruction process, we developed an uncertainty-aware active sensing strategy with diversity-aware selection of measurement locations, thereby connecting reconstruction and sensing in a closed loop. Experiments on simulated electromagnetic maps and a high-fidelity simulated satellite monitoring scenario demonstrated that GSC consistently improves reconstruction quality under sparse, noisy, and low-bit quantized measurements, and sensing efficiency over representative baselines. Future work will extend this framework to spatiotemporal spectrum cartography in dynamic radio environments and explore distributed multi-agent collaborative sensing under communication and coordination constraints.

\appendices

\section{}
\label{app:linear_posterior_mean}

This appendix proves the closed-form posterior mean update in \eqref{eq:linear_posterior_mean}.

Following the same approximation spirit as DMPS~\cite{meng2022diffusion}, we approximate the diffusion-induced conditional prior by a local Gaussian centered at the DDPM estimate $\hat{\boldsymbol{x}}_{0|t}$:
\begin{equation}
p(\boldsymbol{x}_0 | \boldsymbol{x}_t)
\approx
\mathcal{N}
\left(
\hat{\boldsymbol{x}}_{0|t},
\gamma_t^2\mathbf{I}
\right),
\label{eq:gaussian_prior_approx}
\end{equation}
where
$
\gamma_t=\sqrt{\frac{1-\bar{\alpha}_t}{\bar{\alpha}_t}}.
\label{eq:gamma_t_definition}
$

Under the linear measurement model in \eqref{eq:linear-obs-model}, Bayes' rule gives
\begin{equation}
p(\boldsymbol{x}_0 | \boldsymbol{x}_t,\boldsymbol{y})
=
\frac{
p(\boldsymbol{y}|\boldsymbol{x}_0,\boldsymbol{x}_t)\,
p(\boldsymbol{x}_0|\boldsymbol{x}_t)
}{
p(\boldsymbol{y}|\boldsymbol{x}_t)
}.
\label{eq:bayes_cond_xt}
\end{equation}
Since $\boldsymbol{y}$ is conditionally independent of $\boldsymbol{x}_t$ given $\boldsymbol{x}_0$, the posterior satisfies
\begin{equation}
p(\boldsymbol{x}_0 | \boldsymbol{x}_t,\boldsymbol{y})
\propto
p(\boldsymbol{x}_0|\boldsymbol{x}_t)\,p(\boldsymbol{y}|\boldsymbol{x}_0).
\label{eq:posterior_propto_measure}
\end{equation}
The likelihood induced by \eqref{eq:linear-obs-model} is
\begin{equation}
p(\boldsymbol{y}|\boldsymbol{x}_0)
=
\mathcal{N}(\boldsymbol{y};\mathbf{H}\boldsymbol{x}_0,\sigma_y^2\mathbf{I}).
\label{eq:linear_likelihood_app}
\end{equation}
Combining \eqref{eq:gaussian_prior_approx} and \eqref{eq:linear_likelihood_app} yields a standard linear-Gaussian posterior. Its posterior mean is
\begin{equation}
\hat{\boldsymbol{x}}_{0|t,\boldsymbol{y}}
=
\hat{\boldsymbol{x}}_{0|t}
+
\mathbf{K}_t
\left(
\boldsymbol{y}-\mathbf{H}\hat{\boldsymbol{x}}_{0|t}
\right),
\label{eq:linear_kalman_form_app}
\end{equation}
with
\begin{equation}
\mathbf{K}_t
=
\gamma_t^2\mathbf{H}^{\mathsf T}
\left(
\gamma_t^2\mathbf{H}\mathbf{H}^{\mathsf T}
+
\sigma_y^2\mathbf{I}
\right)^{-1}.
\label{eq:linear_gain_app}
\end{equation}
Since $\mathbf{H}$ is a diagonal binary masking matrix, one has
\(
\mathbf{H}^{\mathsf T}=\mathbf{H}
\)
and
\(
\mathbf{H}\mathbf{H}^{\mathsf T}=\mathbf{H}^2=\mathbf{H},
\)
which reduces \eqref{eq:linear_kalman_form_app} to \eqref{eq:linear_posterior_mean}.

\section{}
\label{app:quant_posterior_mean}

This appendix proves the closed-form posterior mean update in \eqref{eq:quant_posterior_mean}.

We now consider the quantized measurement model in \eqref{eq:quant_model}. As in the linear case, the posterior mean is determined by the same prior term in \eqref{eq:gaussian_prior_approx}, but the likelihood becomes nonlinear and non-Gaussian under quantization.

Let $s_t$, $a_i$, and $b_i$ be defined as in \eqref{eq:quant_st} and \eqref{eq:quant_ab}. 
Under the local Gaussian approximation in \eqref{eq:gaussian_prior_approx}, we define
$
\mu_i \triangleq (\hat{\boldsymbol{x}}_{0|t})_i
$
as the mean of the $i$-th component,
so that the corresponding scalar conditional prior for $x_{0,i}$ is
\begin{equation}
p(x_{0,i}|\boldsymbol{x}_t)
=
\mathcal{N}(x_{0,i};\mu_i,\gamma_t^2).
\label{eq:quant_prior_i_app}
\end{equation}
Next, we introduce the latent pre-quantization variable
\begin{equation}
r_i=x_{0,i}+n_i,
\quad
n_i\sim\mathcal{N}(0,\sigma_y^2).
\label{eq:quant_latent_ri_app}
\end{equation}
The measurement $y_i$ corresponds to the interval event 
$l_i<r_i\le u_i.$
Accordingly, the scalar likelihood can be written as
\begin{equation}
p(y_i|x_{0,i})
=
\Phi\!\left(\frac{u_i-x_{0,i}}{\sigma_y}\right)
-
\Phi\!\left(\frac{l_i-x_{0,i}}{\sigma_y}\right).
\label{eq:quant_likelihood_i_app}
\end{equation}
Combining \eqref{eq:quant_prior_i_app} and \eqref{eq:quant_likelihood_i_app}, the posterior density is proportional to a Gaussian prior modulated by an interval-probability likelihood, which induces a truncated-Gaussian structure. Let
\begin{equation}
C_i=\Phi(b_i)-\Phi(a_i)
\label{eq:quant_normalizer_app}
\end{equation}
denote the corresponding normalizing constant. Applying the standard derivative identity for Gaussian exponential families yields
\begin{equation}
\mathbb{E}[x_{0,i}|\boldsymbol{x}_t,y_i]
=
\mu_i+\gamma_t^2\frac{\partial \log C_i}{\partial \mu_i}.
\label{eq:quant_derivative_identity_app}
\end{equation}
Moreover,
\begin{equation}
\frac{\partial \log C_i}{\partial \mu_i}
=
\frac{1}{s_t}\,
\frac{\phi(a_i)-\phi(b_i)}{\Phi(b_i)-\Phi(a_i)}
=
\frac{1}{s_t}\Delta_i,
\label{eq:quant_logC_derivative_app}
\end{equation}
where $\Delta_i$ is defined in \eqref{eq:quant_delta}. Substituting \eqref{eq:quant_logC_derivative_app} into \eqref{eq:quant_derivative_identity_app} gives
\begin{equation}
\mathbb{E}[x_{0,i}|\boldsymbol{x}_t,y_i]
=
\mu_i+\frac{\gamma_t^2}{s_t}\Delta_i.
\label{eq:quant_posterior_i_app}
\end{equation}
For unmeasured coordinates $i\notin\Omega$, the likelihood is uninformative with respect to $x_{0,i}$, so the posterior mean remains equal to the prior mean $\mu_i$. Stacking the coordinate-wise posterior means over all grid entries yields \eqref{eq:quant_posterior_mean}.

\section*{Acknowledgment}
The authors used ChatGPT (OpenAI) for limited language polishing and wording refinement during the preparation of this manuscript. No AI tool was used to generate the core scientific ideas, methodology, experimental results, or conclusions. All manuscript content was carefully reviewed and verified by the authors.

\bibliography{ref}

@article{zymnis2009compressed,
  title={Compressed sensing with quantized measurements},
  author={Zymnis, Argyrios and Boyd, Stephen and Candes, Emmanuel},
  journal={IEEE Signal Processing Letters},
  volume={17},
  number={2},
  pages={149--152},
  year={2009},
  publisher={IEEE}
}

@article{2015ITCCN...1..406D,
       author = {{Debroy}, Saptarshi and {Bhattacharjee}, Shameek and {Chatterjee}, Mainak},
        title = "{Spectrum Map and Its Application in Resource Management in Cognitive Radio Networks}",
      journal = {IEEE Transactions on Cognitive Communications and Networking},
         year = 2015,
        month = jan,
       volume = {1},
       number = {4},
        pages = {406-419},
          doi = {10.1109/TCCN.2016.2517001},
       adsurl = {https://ui.adsabs.harvard.edu/abs/2015ITCCN...1..406D}
}

@inproceedings{denkovski2012reliability,
  title={Reliability of a radio environment map: Case of spatial interpolation techniques},
  author={Denkovski, Daniel and Atanasovski, Vladimir and Gavrilovska, Liljana and Riihij{\"a}rvi, Janne and M{\"a}h{\"o}nen, Petri},
  booktitle={2012 7th international ICST conference on cognitive radio oriented wireless networks and communications (CROWNCOM)},
  pages={248--253},
  year={2012},
  organization={IEEE}
}

@article{bazerque2011group,
  title={Group-lasso on splines for spectrum cartography},
  author={Bazerque, Juan Andr{\'e}s and Mateos, Gonzalo and Giannakis, Georgios B},
  journal={IEEE Transactions on Signal Processing},
  volume={59},
  number={10},
  pages={4648--4663},
  year={2011},
  publisher={IEEE}
}

@INPROCEEDINGS{7952827,
  author={Hamid, Mohamed and Beferull-Lozano, Baltasar},
  booktitle={2017 IEEE International Conference on Acoustics, Speech and Signal Processing (ICASSP)}, 
  title={Non-parametric spectrum cartography using adaptive radial basis functions}, 
  year={2017},
  volume={},
  number={},
  pages={3599-3603},
  doi={10.1109/ICASSP.2017.7952827}}

@INPROCEEDINGS{6362597,
  author={Boccolini, Gabriele and Hernández-Peñaloza, Gustavo and Beferull-Lozano, Baltasar},
  booktitle={2012 IEEE 23rd International Symposium on Personal, Indoor and Mobile Radio Communications - (PIMRC)}, 
  title={Wireless sensor network for Spectrum Cartography based on Kriging interpolation}, 
  year={2012},
  volume={},
  number={},
  pages={1565-1570},
  doi={10.1109/PIMRC.2012.6362597}}

@inproceedings{jayawickrama2013improved,
  title={Improved performance of spectrum cartography based on compressive sensing in cognitive radio networks},
  author={Jayawickrama, Beeshanga Abewardana and Dutkiewicz, Eryk and Oppermann, Ian and Fang, Gengfa and Ding, Jie},
  booktitle={2013 IEEE International Conference on Communications (ICC)},
  pages={5657--5661},
  year={2013},
  organization={IEEE}
}

@ARTICLE{5352337,
  author={Bazerque, Juan Andrés and Giannakis, Georgios B.},
  journal={IEEE Transactions on Signal Processing}, 
  title={Distributed Spectrum Sensing for Cognitive Radio Networks by Exploiting Sparsity}, 
  year={2010},
  volume={58},
  number={3},
  pages={1847-1862},
  doi={10.1109/TSP.2009.2038417}}

@article{zhang2020spectrum,
  title={Spectrum cartography via coupled block-term tensor decomposition},
  author={Zhang, Guoyong and Fu, Xiao and Wang, Jun and Zhao, Xi-Le and Hong, Mingyi},
  journal={IEEE Transactions on Signal Processing},
  volume={68},
  pages={3660--3675},
  year={2020},
  publisher={IEEE}
}

@article{marin2018compressive,
  title={Compressive multispectral spectrum sensing for spectrum cartography},
  author={Mar{\'\i}n Alfonso, Jeison and Mart{\'\i}nez Torre, Jose Ignacio and Arguello Fuentes, Henry and Agudelo, Leonardo Betancur},
  journal={Sensors},
  volume={18},
  number={2},
  pages={387},
  year={2018},
  publisher={MDPI}
}

@INPROCEEDINGS{7178572,
  author={Romero, Daniel and Kim, Seung-Jun and López-Valcarce, Roberto and Giannakis, Georgios B.},
  booktitle={2015 IEEE International Conference on Acoustics, Speech and Signal Processing (ICASSP)}, 
  title={Spectrum cartography using quantized observations}, 
  year={2015},
  volume={},
  number={},
  pages={3252-3256},
  doi={10.1109/ICASSP.2015.7178572}}

@article{Xu2021,
  author = {Y. Xu and W. Zi and J. Song and R. Shao and H. Chen},
  title = {Spatiotemporal correlation analysis and visualization of electromagnetic intensity based on multi-site and multi-time attention mechanism},
  journal = {Journal of Nanjing University (Natural Sciences)},
  volume = {57},
  number = {5},
  pages = {838-846},
  year = {2021},
  doi = {10.13232/j.cnki.jnju.2021.05.014}
}

@article{ho2020denoising,
  title={Denoising diffusion probabilistic models},
  author={Ho, Jonathan and Jain, Ajay and Abbeel, Pieter},
  journal={Advances in neural information processing systems},
  volume={33},
  pages={6840--6851},
  year={2020}
}

@inproceedings{nichol2021improved,
  title={Improved denoising diffusion probabilistic models},
  author={Nichol, Alexander Quinn and Dhariwal, Prafulla},
  booktitle={International conference on machine learning},
  pages={8162--8171},
  year={2021},
  organization={PMLR}
}

@article{song2020denoising,
  title={Denoising diffusion implicit models},
  author={Song, Jiaming and Meng, Chenlin and Ermon, Stefano},
  journal={arXiv preprint arXiv:2010.02502},
  year={2020}
}

@inproceedings{
chung2023diffusion,
title={Diffusion Posterior Sampling for General Noisy Inverse Problems},
author={Hyungjin Chung and Jeongsol Kim and Michael Thompson Mccann and Marc Louis Klasky and Jong Chul Ye},
booktitle={The Eleventh International Conference on Learning Representations },
year={2023}
}

@inproceedings{dou2024diffusion,
  title={Diffusion posterior sampling for linear inverse problem solving: A filtering perspective},
  author={Dou, Zehao and Song, Yang},
  booktitle={The Twelfth International Conference on Learning Representations},
  year={2024}
}

@INPROCEEDINGS{4699911,
  author={Alaya-Feki, Afef Ben Hadj and Jemaa, Sana Ben and Sayrac, Berna and Houze, Paul and Moulines, Eric},
  booktitle={2008 IEEE 19th International Symposium on Personal, Indoor and Mobile Radio Communications}, 
  title={Informed spectrum usage in cognitive radio networks: Interference cartography}, 
  year={2008},
  volume={},
  number={},
  pages={1-5},
  doi={10.1109/PIMRC.2008.4699911}}

@article{liu2022flow,
  title={Flow straight and fast: Learning to generate and transfer data with rectified flow},
  author={Liu, Xingchao and Gong, Chengyue and Liu, Qiang},
  journal={arXiv preprint arXiv:2209.03003},
  year={2022}
}

@ARTICLE{10764739,
  author={Wang, Xiucheng and Tao, Keda and Cheng, Nan and Yin, Zhisheng and Li, Zan and Zhang, Yuan and Shen, Xuemin},
  journal={IEEE Transactions on Cognitive Communications and Networking}, 
  title={RadioDiff: An Effective Generative Diffusion Model for Sampling-Free Dynamic Radio Map Construction}, 
  year={2025},
  volume={11},
  number={2},
  pages={738-750},
  doi={10.1109/TCCN.2024.3504489}}

@article{lipman2022flow,
  title={Flow matching for generative modeling},
  author={Lipman, Yaron and Chen, Ricky TQ and Ben-Hamu, Heli and Nickel, Maximilian and Le, Matt},
  journal={arXiv preprint arXiv:2210.02747},
  year={2022}
}

@article{shrestha2022deep,
  title={Deep spectrum cartography: Completing radio map tensors using learned neural models},
  author={Shrestha, Sagar and Fu, Xiao and Hong, Mingyi},
  journal={IEEE Transactions on Signal Processing},
  volume={70},
  pages={1170--1184},
  year={2022},
  publisher={IEEE}
}

@ARTICLE{11127065,
  author={Timilsina, Subash and Shrestha, Sagar and Cheng, Lei and Fu, Xiao},
  journal={IEEE Signal Processing Letters}, 
  title={Domain-Factored Untrained Deep Prior for Spectrum Cartography}, 
  year={2025},
  volume={32},
  number={},
  pages={3440-3444},
  doi={10.1109/LSP.2025.3599714}}

@inproceedings{zhang2018unreasonable,
  title={The unreasonable effectiveness of deep features as a perceptual metric},
  author={Zhang, Richard and Isola, Phillip and Efros, Alexei A and Shechtman, Eli and Wang, Oliver},
  booktitle={Proceedings of the IEEE conference on computer vision and pattern recognition},
  pages={586--595},
  year={2018}
}

@article{wang2022zero,
  title={Zero-shot image restoration using denoising diffusion null-space model},
  author={Wang, Yinhuai and Yu, Jiwen and Zhang, Jian},
  journal={arXiv preprint arXiv:2212.00490},
  year={2022}
}

@inproceedings{song2023pseudoinverse,
  title={Pseudoinverse-guided diffusion models for inverse problems},
  author={Song, Jiaming and Vahdat, Arash and Mardani, Morteza and Kautz, Jan},
  booktitle={International Conference on Learning Representations},
  year={2023}
}

@article{reddy2022spectrum,
  title={Spectrum cartography techniques, challenges, opportunities, and applications: A survey},
  author={Reddy, Yeduri Sreenivasa and Kumar, Abhinav and Pandey, Om Jee and Cenkeramaddi, Linga Reddy},
  journal={Pervasive and Mobile Computing},
  volume={79},
  pages={101511},
  year={2022},
  publisher={Elsevier}
}

@ARTICLE{10906396,
  author={Pan, Zhen and Bangning, Zhang and Heng, Wang and Wenfeng, Ma and Daoxing, Guo},
  journal={China Communications}, 
  title={SC-GAN: A spectrum cartography with satellite Internet based on Pix2Pix generative adversarial network}, 
  year={2025},
  volume={22},
  number={2},
  pages={47-61},
  doi={10.23919/JCC.fa.2024-0269.202502}}

@article{song2020score,
  title={Score-based generative modeling through stochastic differential equations},
  author={Song, Yang and Sohl-Dickstein, Jascha and Kingma, Diederik P and Kumar, Abhishek and Ermon, Stefano and Poole, Ben},
  journal={arXiv preprint arXiv:2011.13456},
  year={2020}
}

@article{kawar2022denoising,
  title={Denoising diffusion restoration models},
  author={Kawar, Bahjat and Elad, Michael and Ermon, Stefano and Song, Jiaming},
  journal={Advances in neural information processing systems},
  volume={35},
  pages={23593--23606},
  year={2022}
}

@inproceedings{meng2022diffusion,
  author       = {Xiangming Meng and
                  Yoshiyuki Kabashima},
  editor       = {Vu Nguyen and
                  Hsuan{-}Tien Lin},
  title        = {Diffusion Model Based Posterior Sampling for Noisy Linear Inverse
                  Problems},
  booktitle    = {Asian Conference on Machine Learning, 5-8 December 2024, Hanoi, Vietnam},
  series       = {Proceedings of Machine Learning Research},
  volume       = {260},
  pages        = {623--638},
  publisher    = {{PMLR}},
  year         = {2024},
}

@article{song2019generative,
  title={Generative modeling by estimating gradients of the data distribution},
  author={Song, Yang and Ermon, Stefano},
  journal={Advances in neural information processing systems},
  volume={32},
  year={2019}
}

@inproceedings{zhang2025improving,
  title={Improving diffusion inverse problem solving with decoupled noise annealing},
  author={Zhang, Bingliang and Chu, Wenda and Berner, Julius and Meng, Chenlin and Anandkumar, Anima and Song, Yang},
  booktitle={Proceedings of the Computer Vision and Pattern Recognition Conference},
  pages={20895--20905},
  year={2025}
}

@inproceedings{meng2024qcs,
  title={QCS-SGM+: Improved Quantized Compressed Sensing with Score-Based Generative Models},
  author={Meng, Xiangming and Kabashima, Yoshiyuki}, 
  booktitle={Proceedings of the AAAI Conference on Artificial Intelligence},
  volume={38},
  number={13},
  pages={14341--14349},
  year={2024}
}

@inproceedings{meng2022quantized,
  title={Quantized Compressed Sensing with Score-Based Generative Models},
  author={Meng, Xiangming and Kabashima, Yoshiyuki},
  booktitle={International Conference on Learning Representations},
  year={2023}
}

@article{article1,
author = {Huang, Yang and Cui, Haoyu and Hou, Yuqi and Hao, Caiyong and Wang, Wei and Zhu, Qiuming and Li, Jie and Wu, Qihui and Wang, Jiabo},
year = {2024},
month = {02},
pages = {},
title = {Space-Based Electromagnetic Spectrum Sensing and Situation Awareness},
volume = {4},
journal = {Space: Science \& Technology},
doi = {10.34133/space.0109}
}

@ARTICLE{9520322,
  author={Hao, Caiyong and Wan, Xianrong and Feng, Daquan and Feng, Zhiyong and Xia, Xiang-Gen},
  journal={IEEE Network}, 
  title={Satellite-Based Radio Spectrum Monitoring: Architecture, Applications, and Challenges}, 
  year={2021},
  volume={35},
  number={4},
  pages={20-27},
  doi={10.1109/MNET.011.2100015}}

@inproceedings{macqueen1967kmeans,
  author    = {J. MacQueen},
  title     = {Some Methods for Classification and Analysis of Multivariate Observations},
  booktitle = {Proceedings of the Fifth Berkeley Symposium on Mathematical Statistics and Probability},
  volume    = {1},
  pages     = {281--297},
  year      = {1967},
  publisher = {University of California Press}
}

@book{bishop2006prml,
  author    = {Christopher M. Bishop},
  title     = {Pattern Recognition and Machine Learning},
  year      = {2006},
  publisher = {Springer},
  address   = {New York}
}

@book{murphy2012mlpp,
  author    = {Kevin P. Murphy},
  title     = {Machine Learning: A Probabilistic Perspective},
  year      = {2012},
  publisher = {The MIT Press},
  address   = {Cambridge, MA}
}

@ARTICLE{8648450,
  author={Bi, Suzhi and Lyu, Jiangbin and Ding, Zhi and Zhang, Rui},
  journal={IEEE Wireless Communications}, 
  title={Engineering Radio Maps for Wireless Resource Management}, 
  year={2019},
  volume={26},
  number={2},
  pages={133-141},
  doi={10.1109/MWC.2019.1800146}}
\bibliographystyle{IEEEtran}

\end{document}